\documentclass[10pt]{article}  
 \usepackage{amsmath,amssymb} 

\usepackage{fullpage}
\usepackage[]{graphicx}
\graphicspath{{graphics/}} 
\usepackage{setspace}
\usepackage{tocloft}
\usepackage{color}
\usepackage[linktocpage]{hyperref}
\usepackage{cite}
\usepackage{wrapfig}
\usepackage{lscape}

\def\eps{\varepsilon}
\newcommand{\fraz}{\displaystyle\frac}
\def\##1{{\bf #1}}
\def\=#1{\underline{\underline #1}}
\def\ux{\#u_x}
\def\uy{\#u_y}
\def\uz{\#u_z}
\def\Pa{{\cal P}_a}
\def\Pb{{\cal P}_b}
\def\ped0{_{\scriptscriptstyle 0}}
\def\ko{k\ped0}
\def\tq{\tilde{q}}
\def\vph{v_{ph}}
\def\propdist{\Delta_{prop}}
\def\fref#1{Figure~\ref{#1}}

\def\sref#1{Section~\ref{#1}}
\def\tref#1{Table~\ref{#1}}
\def\eref#1{Eq.~\eqref{#1}}

\def\tond#1{\left(#1\right)}
\def\quadr#1{\left[#1\right]}
\def\graff#1{\left\{#1\right\}} 
\newcommand*\diff{\mathop{}\!\mathrm{d}}

\cftpagenumbersoff{figure}
\cftpagenumbersoff{table} 
\begin{document} 


\begin{center}
\textbf{Compound surface-plasmon-polariton waves guided by a thin metal layer sandwiched between a homogeneous isotropic dielectric material and a periodically multilayered isotropic dielectric material}\\

\textit{Francesco Chiadini$^{a}$, Vincenzo Fiumara$^{b}$, Antonio 
Scaglione$^{a}$, and Akhlesh Lakhtakia$^{c}$}\\

{$^{a}$University of 
Salerno, Department of Industrial Engineering, \\ via Giovanni Paolo II, 132 - Fisciano (SA), 84084, Italy\\
$^{b}$University of 
Basilicata, School of Engineering, \\ Viale 
dell'Ateneo Lucano 10, 85100 Potenza, Italy\\
$^{c}$The Pennsylvania State University, Department of Engineering 
Science and Mechanics, \\ Nanoengineered Metamaterials Group, University Park, PA 
16802--6812, USA}
\end{center}

\begin{abstract}
Multiple $p$- and $s$-polarized compound surface plasmon-polariton (SPP)  waves at a fixed frequency can be guided by a structure consisting of  a metal layer sandwiched between a homogeneous isotropic dielectric (HID) material and a periodic multilayered isotropic dielectric (PMLID) material. 
For any thickness of the metal layer,  at least one compound SPP wave must exist. It possesses the $p$-polarization state, is strongly bound to the metal/HID interface when the metal thickness is large but to both  metal/dielectric interfaces when the metal thickness is small. When the metal layer vanishes, this compound SPP wave transmutes into a Tamm wave.  
Additional compound SPP waves exist, depending on the thickness of the metal layer, the relative permittivity of the
HID material, and the period and the composition of the PMLID material.
Some  of these are $p$ polarized, the others being $s$ polarized. All of them differ in phase speed, attenuation rate, and field profile, even though all are excitable at the same frequency.  The multiplicity and the dependence of the number of compound SPP waves on the relative permittivity of the HID material when the metal layer is thin could be useful for optical sensing applications.

\end{abstract}

 
\section{Introduction}
\label{sect:intro} 
The electromagnetic fields of a
 surface-plasmon-polariton (SPP) wave  guided by  the planar interface of a homogenous metal and a homogeneous isotropic dielectric material  decay  along the direction normal to the  interface in both materials~\cite{Pitarke}. The SPP wave is the classical counterpart of a train of quantum quasi-particles formed jointly by plasmons and polaritons bound 
 to the metal/dielectric interface. The SPP wave is $p$-polarized. Only one SPP wave can be guided by the interface at a specific optical frequency. Due to the field being bound tightly to the metal/dielectric interface, the SPP wave is strongly sensitive to the constitutive parameters of the materials on either side of the interface, leading to optical sensing applications~\cite{AZLreview,ZZX,Couture}. For the same reason,
researchers have also focused on  surface-plasmon-resonance   microscopes
with subwavelength resolution to obtain images of objects with very low contrast, without the use of dyes
or markers \cite{Berguiga,Stabler}. Applications of SPP waves are on the horizon in the area of communications
 as well \cite{Sekhon}.

The excitation of a multiplicity of SPP waves at a specific frequency is very appealing therefore. This multiplicity is possible if  the isotropic  dielectric partner is  periodically nonhomogeneous in the direction normal to the interface,
which has been established both theoretically  \cite{PLreview,AkhBook} and experimentally \cite{Hallnano,Liujnp}.
Even though all SPP waves are excited at the same frequency, they will differ in one or more of the following attributes:
polarization state, phase speed, attenuation rate, and field profile.  

In the present paper, we theoretically formulate and analyze a boundary-value problem that involves two parallel metal/dielectric interfaces in a bid to coalesce the SPP waves guided separately by each interface into compound SPP waves. For simplicity, one of the two dielectric materials is taken to be homogeneous while the other dielectric material is
periodically multilayered normal to the interfaces. Both dielectric materials are isotropic as also is the metal separating them.

\begin{figure}
 \begin{center}
\includegraphics[width=0.5\linewidth]{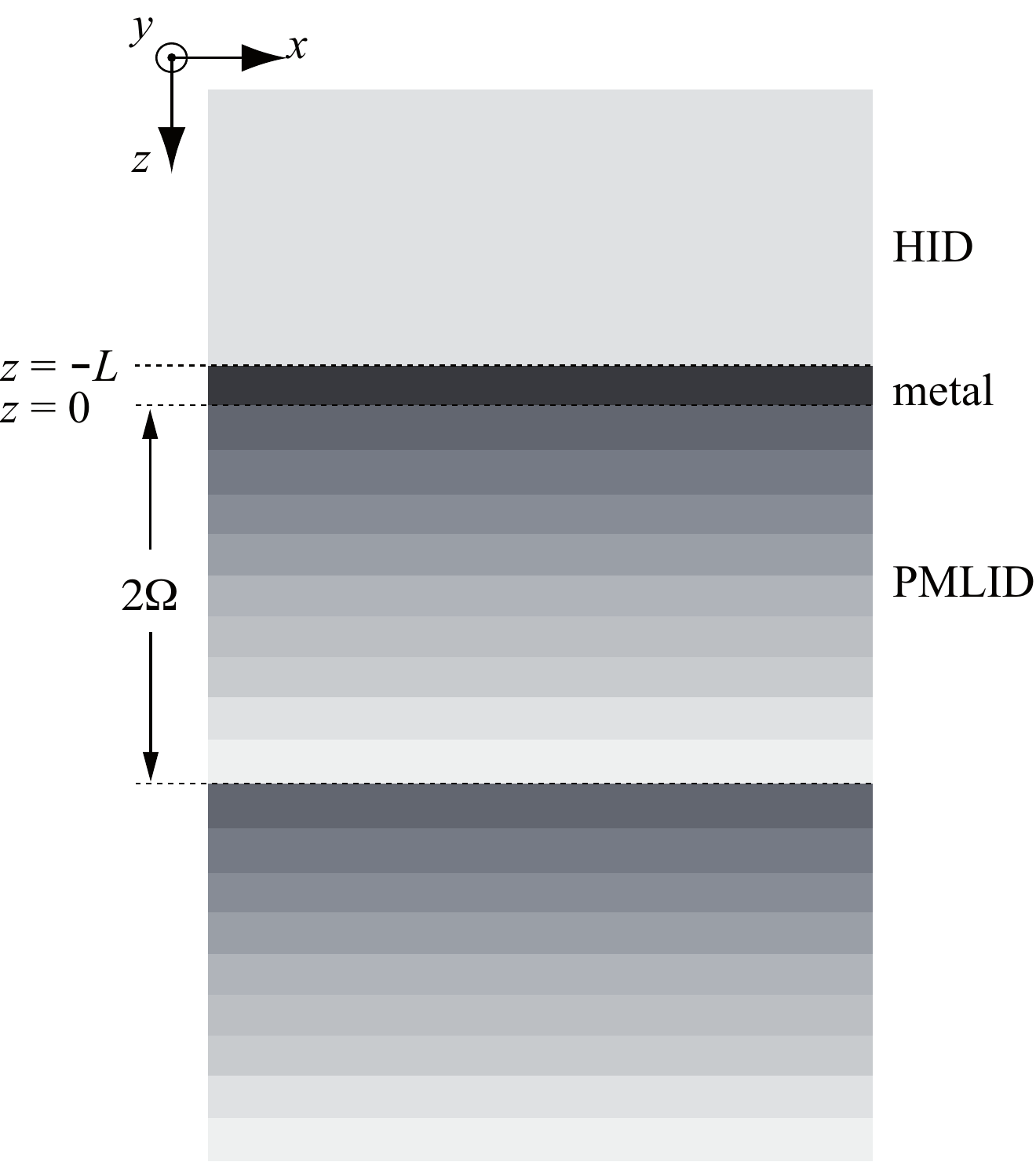}
 \end{center}
\caption{Schematic of the boundary-value problem. A metal layer of thickness $L$ separates a half space
occupied by a HID material and another half space occupied by a PMLID material of period $2\Omega$.
}
\label{fig:schem}
 \end{figure}

The plan of this paper is as follows. \sref{sec:theo} presents the formulation of the boundary-value problem, wherein
one half space is occupied by a homogeneous isotropic dielectric (HID) material, another half space is occupied by
a periodic multilayered isotropic dielectric (PMLID) material, and the two half spaces are separated by a thin metallic layer,
as shown in \fref{fig:schem}. In this structure, compound SPP waves of the $p$-polarization state can not interact with compound SPP waves
of the $s$-polarization state, thereby splitting the boundary-value problem into two autonomous sub-problems. Numerical results showing the compounding of the SPP waves guided by the metal/HID and metal/PMLID interfaces are presented
 in relation to the thickness of the metal layer   in \sref{sec:numres}. Concluding remarks are presented in \sref{sec:concl}. 
 
 An $\exp\tond{-i\omega t}$ dependence on time $t$ is implicit, with $\omega$ denoting the angular frequency and $i=\sqrt{-1}$. Furthermore, $\ko =\omega \sqrt{\eps\ped0\mu\ped0}$, $c\ped0=\omega/\ko$, $\eta\ped0=\sqrt{\mu\ped0/\eps\ped0}$ and $\lambda\ped0=2\pi/\ko$, respectively, represent the   wavenumber,   phase speed, characteristic impedance, and   wavelength
 in free space, with $\mu\ped0$ as the permeability and $\eps\ped0$ as the permittivity of free space. Vectors are in boldface, column vectors are in boldface and enclosed within square brackets, and matrices are twice underlined and square bracketed. The asterisk denotes the complex conjugate. Cartesian unit vectors are denoted by $\ux$, $\uy$, and $\uz$.
 
\section{Theoretical framework}\label{sec:theo}

The schematic of the boundary-value problem is shown
in \fref{fig:schem}.
The half space $z < -L$ is occupied by a HID material with real-valued relative permittivity $\eps_d>0$.   
A metal layer of thickness $L$ and complex-valued
relative permittivity $\eps_m$ separates  the HID material from the PMLID material which occupies
the half space $z>0$. The unit cell of the PMLID material consists of $N$ layers
with real-valued relative permittivity $\eps_j>0$ and   thickness $d_j$ $(j=1,\ldots,N)$, giving the period $2\Omega=d_1+d_2+\ldots+d_N$. 

Without loss of generality, we consider the direction of propagation of the compound SPP wave to be parallel to the $x$ axis. Due to the isotropy of all materials, we can consider $p$- and $s$-polarized compound SPP waves guided by the
structure in \fref{fig:schem} separately from each other.

\subsection{$p$-polarized compound SPP waves}\label{sec:2p}
The electric and magnetic field phasors of a $p$-polarized compound SPP wave can be represented everywhere as
\begin{equation}
\left.\begin{array}{l}
\#E(\#r)= \quadr{e_x(z)\ux+e_z(z)\uz}\exp({iqx})\\[5pt]
\#H(\#r)= h_y(z) \uy\exp({iqx})
\end{array}\right\}\,, \quad z \in(-\infty,\infty)\,,
\label{eq:EHpmlid}
\end{equation}
where $q$ is the complex-valued wavenumber. The functions $e_x(z)$, $e_z(z)$, and $h_y(z)$ are determined piecewise as follows.

In the j$^{th}$ layer of the PMLID material,
the Faraday and the Amp\'ere--Maxwell equations yield
\begin{equation}
e_z(z) = -\fraz{q}{\omega\eps\ped0\eps_j}h_y(z)\,
\label{eq:Ez}
\end{equation}
and
\begin{equation}
\fraz{\diff }{\diff z}\quadr{\#f^{\tond{p}}\tond{z}}=i\quadr{\=P_j^{\tond{p}}} \cdot\quadr{\#f^{\tond{p}}\tond{z}}\,,
\label{eq:ODE_PMLID}
\end{equation}
where  the column vector
\begin{equation}
\quadr{\#f^{\tond{p}}\tond{z}}=
\begin{bmatrix}
e_x(z)\\[5pt]
h_y(z)
\end{bmatrix}\,
\label{eq:fp}
\end{equation}
and the matrix 
\begin{equation}
\quadr{\=P_j^{\tond{p}}}=
\begin{bmatrix}
 0 & \omega\mu\ped0-\fraz{q^2}{\omega\eps\ped0\eps_j}\\[10pt]
\omega\eps\ped0\eps_j & 0
\end{bmatrix}\,.
\end{equation}
Since the matrix $\quadr{\=P_j^{\tond{p}}}$ is independent of $z$ inside the j$^{th}$ layer, the solution
of \eref{eq:ODE_PMLID} is straightforward \cite{AkhBook,Hochstadt}. Most importantly, we get
\cite{FHBML}
\begin{equation}
\quadr{\#f^{\tond{p}}\tond{2\Omega^-}} = \quadr{\=Q^{\tond{p}}}\cdot\quadr{\#f^{\tond{p}}\tond{0^+}}\,,
\label{eq:metalP}
\end{equation}
where
\begin{equation}
\quadr{\=Q^{\tond{p}}}=\exp\graff{i\quadr{\=P_N^{\tond{p}}}d_N}
\cdot\exp\graff{i\quadr{\=P_{N-1}^{\tond{p}}}d_{N-1}}\cdot\ldots\cdot
\exp\graff{i\quadr{\=P_2^{\tond{p}}}d_2}\cdot\exp\graff{i\quadr{\=P_1^{\tond{p}}}d_1}\,.
\label{eq:Qp}
\end{equation}

Let the two eigenvalues of $\quadr{\=Q^{\tond{p}}}$ be denoted by $\sigma_1^{\tond{p}}$ and
$\sigma_2^{\tond{p}}$, and the respective eigenvectors by $\quadr{\#t_1^{\tond{p}}}$
and $\quadr{\#t_2^{\tond{p}}}$. One of the two eigenvalues represents a field decaying as $z\to\infty$,
the other a field decaying as $z\to-\infty$. Since the fields of the compound SPP wave must decay in the
half space occupied by the PMLID material \cite{AkhBook,FHBML}
\begin{equation}
\quadr{\#f^{\tond{p}}\tond{0^+}}= c_p \quadr{\#t_1^{\tond{p}}}\,,
\label{eq:f_0p}
\end{equation}
provided that the eigenvalues are labeled so that
${\rm Re}\left(\ln\sigma_1^{\tond{p}}\right) <0$ and, \textit{ipso facto}, ${\rm Re}\left(\ln\sigma_2^{\tond{p}}\right) >0$. The coefficient $c_p$ is as yet unknown.
 
In the half space $z<-L$ occupied by the HID material, field representation is of the textbook variety; thus,
with $b_p$ as an unknown coefficient we have
\begin{equation}
\left.\begin{array}{c}
\#E(\#r) = b_p\left(\fraz{-\alpha_d\ux+q\uz}{\ko n_d}\right)\exp(iqx)\exp\left[i\alpha_d(z+L)\right]\\[8pt]
\#H(\#r) = -b_p\uy\fraz{n_d}{\eta\ped0}\exp(iqx)\exp\left[i\alpha_d(z+L)\right]
\end{array}\right\}\,,\quad z < -L\,,
\end{equation}
where $n_d = +\sqrt{\eps_d}$, $\alpha_d^2+q^2=\ko^2\eps_d$, and ${\rm Im}\left(\alpha_d\right)<0$. Accordingly,
\begin{equation}
\quadr{\#f^{\tond{p}}\tond{-L^-}}=-b_p
\begin{bmatrix}
\fraz{\alpha_d}{\ko n_d}\\[8pt]
\fraz{n_d}{\eta\ped0}
\end{bmatrix}\,.
\label{eq:f-Lp}
\end{equation}

Inside the metal slab, the Faraday and the Amp\'ere--Maxwell equations yield
\begin{equation}
e_z(z) = -\fraz{q}{\omega\eps\ped0\eps_m}h_y(z)\,,\quad -L<z<0\,,
\end{equation}
and
\begin{equation}
\fraz{\diff }{\diff z}\quadr{\#f^{\tond{p}}\tond{z}}=i\quadr{\=P_{m}^{\tond{p}}} \cdot\quadr{\#f^{\tond{p}}\tond{z}}
\,,\quad -L<z<0\,,
\label{eq:ODE_metal}
\end{equation}
where  the  matrix 
\begin{equation}
\quadr{\=P_m^{\tond{p}}}=
\begin{bmatrix}
 0 & \omega\mu\ped0-\fraz{q^2}{\omega\eps\ped0\eps_m}\\[10pt]
\omega\eps\ped0\eps_m & 0
\end{bmatrix}\,.
\end{equation}
This differential equation yields \cite{Hochstadt}
\begin{equation}
\quadr{\#f^{\tond{p}}\tond{0^-}}=\exp\left\{i
\quadr{\=P_m^{\tond{p}}}L\right\}
\cdot
\quadr{\#f^{\tond{p}}\tond{-L^+}}\,.
\label{eq14}
\end{equation}

Standard electromagnetic boundary conditions mandate the equalities
$\quadr{\#f^{\tond{p}}\tond{0^-}}=\quadr{\#f^{\tond{p}}\tond{0^+}}$
and
$\quadr{\#f^{\tond{p}}\tond{-L^-}}=\quadr{\#f^{\tond{p}}\tond{-L^+}}$.
After enforcing these equalities and
substituting
Eqs.~(\ref{eq:f_0p}) and (\ref{eq:f-Lp})  in Eq.~(\ref{eq14}), we get
\begin{equation}
 c_p \quadr{\#t_1^{\tond{p}}}=
-b_p\exp\left\{i
\quadr{\=P_m^{\tond{p}}}L\right\}
\cdot
\begin{bmatrix}
\fraz{\alpha_d}{\ko n_d}\\[8pt]
\fraz{n_d}{\eta\ped0}
\end{bmatrix}\,,
\end{equation}
which can be recast as
\begin{equation}
\quadr{\=Y^{\tond{p}}(q)}\cdot
\begin{bmatrix}
b_p\\
c_p
\end{bmatrix}=\begin{bmatrix}
0\\
0
\end{bmatrix}\,.
\label{eq:Yp}
\end{equation}
The dispersion equation for $p$-polarized compound SPP  waves then is as follows:
\begin{equation}
{\rm det}\,\quadr{\=Y^{\tond{p}}(q)} = 0\,.
\label{eq:detYp}
\end{equation}
For any solution $q$ of \eref{eq:detYp}, the ratio
$b_p/c_p$ can  be found using  \eref{eq:Yp}.

\subsection{$s$-polarized compound SPP waves}\label{sec:2s}

The electric and magnetic field phasors of an $s$-polarized compound SPP wave can be represented everywhere as
\begin{equation}
\left.\begin{array}{l}
\#E(\#r)=  e_y(z) \uy\exp({iqx})\\[5pt]
\#H(\#r)= \quadr{h_x(z)\ux+h_z(z)\uz}
\exp({iqx})
\end{array}\right\}\,, \quad z \in(-\infty,\infty)\,,
\label{eq:EHpmlid-s}
\end{equation}
where the functions $e_y(z)$, $h_x(z)$, and $h_z(z)$ are determined piecewise as follows.

In the j$^{th}$ layer of the PMLID material,
the Faraday and the Amp\'ere--Maxwell equations yield
\begin{equation}
h_z(z) = \fraz{q}{\omega\mu\ped0}e_y(z)
\label{eq:Hz}
\end{equation}
and
\begin{equation}
\fraz{\diff }{\diff z}\quadr{\#f^{\tond{s}}\tond{z}}=i\quadr{\=P_j^{\tond{s}}} \cdot\quadr{\#f^{\tond{s}}\tond{z}}\,,
\label{eq:ODE_PMLID-s}
\end{equation}
where  the column vector
\begin{equation}
\quadr{\#f^{\tond{s}}\tond{z}}=
\begin{bmatrix}
e_y(z)\\[5pt]
h_x(z)
\end{bmatrix}\,
\label{eq:fs}
\end{equation}
and the matrix 
\begin{equation}
\quadr{\=P_j^{\tond{s}}}=
\begin{bmatrix}
 0 & -\omega\mu\ped0\\[10pt]
-\omega\eps\ped0\eps_j +\fraz{q^2}{\omega\mu\ped0} & 0
\end{bmatrix}\,.
\end{equation}
Following~Sec.~\ref{sec:2p}, we get
\cite{FHBML}
\begin{equation}
\quadr{\#f^{\tond{s}}\tond{2\Omega^-}} = \quadr{\=Q^{\tond{s}}}\cdot\quadr{\#f^{\tond{s}}\tond{0^+}}\,,
\label{eq:metalS}
\end{equation}
where
\begin{equation}
\quadr{\=Q^{\tond{s}}}=\exp\graff{i\quadr{\=P_N^{\tond{s}}}d_N}
\cdot\exp\graff{i\quadr{\=P_{N-1}^{\tond{s}}}d_{N-1}}\cdot\ldots\cdot
\exp\graff{i\quadr{\=P_2^{\tond{s}}}d_2}\cdot\exp\graff{i\quadr{\=P_1^{\tond{s}}}d_1}\,.
\label{eq:Qs}
\end{equation}
Accordingly,
\begin{equation}
\quadr{\#f^{\tond{s}}\tond{0^+}}= c_s \quadr{\#t_1^{\tond{s}}}\,,
\label{eq:f_0s}
\end{equation}
where $c_s$ is an unknown coefficient, and the column vector $\quadr{\#t_1^{\tond{s}}}$
satisfies the condition
\begin{equation}
\quadr{\=Q^{\tond{s}}}\cdot\quadr{\#t_1^{\tond{s}}}=\sigma_1^{\tond{s}} \quadr{\#t_1^{\tond{s}}}
\end{equation}
with ${\rm Re}\left(\ln\sigma_1^{\tond{s}}\right) <0$\,.

In the half space $z<-L$ occupied by the HID material,
with $b_s$ as an unknown coefficient we have
\begin{equation}
\left.\begin{array}{c}
\#E(\#r) = b_s\uy\exp(iqx)\exp\left[i\alpha_d(z+L)\right]\\[5pt]
\#H(\#r) = b_s\fraz{n_d}{\eta\ped0}\left(\fraz{-\alpha_d\ux+q\uz}{\ko n_d}\right)\exp(iqx)\exp\left[i\alpha_d(z+L)\right] 
\end{array}\right\}\,,\quad z < -L\,,
\end{equation}
so that
\begin{equation}
\quadr{\#f^{\tond{s}}\tond{-L^-}}=b_s
\begin{bmatrix}
1\\[5pt]
-\fraz{\alpha_d}{\omega\mu\ped0}
\end{bmatrix}\,.
\label{eq:f-Ls}
\end{equation}
Inside the metal slab, 
\begin{equation}
h_z(z) = \fraz{q}{\omega\mu\ped0}e_y(z)\,,\quad -L<z<0\,,
\label{eq:hz}\end{equation}
and
\begin{equation}
\fraz{\diff }{\diff z}\quadr{\#f^{\tond{s}}\tond{z}}=i\quadr{\=P_{m}^{\tond{s}}} \cdot\quadr{\#f^{\tond{s}}\tond{z}}
\,,\quad -L<z<0\,,
\label{eq:ODE_metal-s}
\end{equation}
where  the  matrix 
\begin{equation}
\quadr{\=P_m^{\tond{s}}}=
\begin{bmatrix}
 0 & -\omega\mu\ped0\\[10pt]
-\omega\eps\ped0\eps_m +\fraz{q^2}{\omega\mu\ped0} & 0
\end{bmatrix}\,.
\end{equation}
This differential equation yields \cite{Hochstadt}
\begin{equation}
\quadr{\#f^{\tond{s}}\tond{0^-}}=\exp\left\{i
\quadr{\=P_m^{\tond{s}}}L\right\}
\cdot
\quadr{\#f^{\tond{s}}\tond{-L^+}}\,.
\label{eq32}
\end{equation}

Satisfaction of standard electromagnetic boundary conditions leads to the relation
\begin{equation}
 c_s \quadr{\#t_1^{\tond{s}}}=
b_s\exp\left\{i
\quadr{\=P_m^{\tond{s}}}L\right\}
\cdot
\begin{bmatrix}
1\\[5pt]
-\fraz{\alpha_d}{\omega\mu\ped0}
\end{bmatrix}\,,
\end{equation}
which can be set up as
\begin{equation}
\quadr{\=Y^{\tond{s}}(q)}\cdot
\begin{bmatrix}
b_s\\
c_s
\end{bmatrix}=\begin{bmatrix}
0\\
0
\end{bmatrix}\,.
\label{eq:Ys}
\end{equation}
The dispersion equation for $s$-polarized compound SPP  waves then is as follows:
\begin{equation}
{\rm det}\,\quadr{\=Y^{\tond{s}}(q)} = 0\,.
\label{eq:detYs}
\end{equation}
For any solution $q$ of \eref{eq:detYs}, the ratio
$b_s/c_s$ can  be determined using  \eref{eq:Ys}.

\section{Numerical results}\label{sec:numres}
Equations~(\ref{eq:detYp}) and (\ref{eq:detYs}) were solved numerically using Mathematica (Version 10) on a Windows XP laptop computer. For all calculations, we fixed  $ \lambda\ped0=500$~nm. The HID material and the metal were taken to be SF11 glass ($ \eps_d =3.2508 $) for the results presented in Secs.~\ref{pSPPres} and \ref{sSPPres},
but $\eps_d\in[1,4]$ was kept variable for the results presented in Sec.~\ref{variableHID}.
The  metal was taken to be  silver ($ \eps_m =-7.4726+i 0.8 $). 
  The PMLID material was taken to have a unit cell comprising $N=9$   lossless dielectric layers of  silicon oxide and silicon nitride in different  ratios~\cite{FHBML}. The sequence of the relative  permittivities of layers in the unit cell of the PMLID material is  shown in \tref{tab:PMLID}. 
While $d_j=75$~nm $\forall j\in[1,N]$, 
 the  thickness $L$ of the silver layer was kept variable from 1 to 200~nm. At the chosen wavelength, the skin depth in silver is $ \delta_m= 1/{\rm Im}\left(\ko\sqrt{\eps_m}\right)=29.07$~nm. 
 
In this section, we present the solutions
$q=\tq{k\ped0}$ of the dispersion equations (\ref{eq:detYp}) and (\ref{eq:detYs}). As our computer code had significant  errors in the computation of the transfer matrix $\quadr{\=Q^{\tond{p}}}$ of the PMLID unit cell for ${\rm Re}(\tq)>3.0$, the data presented in this section are restricted to ${\rm Re}(\tq)\leq3.0$.
We also present  the spatial profiles of the Cartesian components of the time-averaged Poynting vector 
\begin{equation}
\textbf{P}\left(x,z \right)=(1/2){\rm Re}\left[\textbf{E}(x,z) \times \textbf{H}^\ast(x,z)\right]
\end{equation}
for representative solutions. These profiles were computed by setting either $b_p=1$ or $b_s=1$, as appropriate.
 
Let us note here that chosen structure can guide surface waves even in the absence of the metal layer, such waves being called Tamm waves \cite{YYH,YYC,MSM,VGM,PFL}. Tamm waves have not only been experimentally observed \cite{RM1999,Robertson},
 but have also been used for optical sensing applications \cite{SR,KA}. Our theoretical framework  accommodates Tamm waves by the simple expedient of setting $L=0$. The solutions of Eqs.~(\ref{eq:detYp}) and (\ref{eq:detYs}) for Tamm waves 
guided by the HID/PMLID interface when $L=0$ are also provided in this section.

\begin{table}[h]
\caption{\bf Relative permittivity $ \eps_j $ at $ \lambda_0=500 $ nm of the j$^{th}$ dielectric layer in the 
unit cell of  the PMLID material
\label{tab:PMLID}
}
\begin{center} 
\begin{tabular}{r|cccccccccc}
\hline
\hline\hline
\rule[-1ex]{0pt}{3.5ex} $ j $ & $1 $ & $2 $ & $ 3$ & $ 4 $ & $ 5 $ & $ 6 $  &  $ 7 $ & $ 8 $ & $ 9 $\\
\rule[-1ex]{0pt}{3.5ex} $\eps_j $ & $ 4.0720$  &  $ 3.3787$  & $ 3.2176 $ & $ 3.0628 $ & $ 2.7936 $ & $ 2.6488 $ & $ 2.4759 $  & $ 2.3388 $  & $ 2.1986 $ \\
\hline
\end{tabular}
\end{center}
\end{table}

\subsection{$p$-polarized compound SPP waves}\label{pSPPres}
The normalized wavenumbers $\tq$ calculated for $p$-polarized compound SPP waves are    reported in \tref{tab:Psol} for different  thicknesses $ L\in[1,200]$~nm of the metal layer. The table also lists the values of $\tq$ for (a) the sole $p$-polarized SPP wave guided by the metal/HID interface by itself and (b) the four $p$-polarized SPP waves guided by the metal/PMLID interface by itself.

\tref{tab:Psol} shows that as many as five different $p$-polarized compound SPP waves can be guided by
 the HID/metal/PMLID structure for $ L\in[1,200]$~nm.  These waves are organized in five sets labeled $p_1$ to $p_5$. Compound SPP waves
in the sets $p_3$, $p_4$, and $p_5$ do not exist if the thickness $L$ of the metal layer is $\lesssim 40.9$~nm, $44.5$~nm, and $46.8$~nm, respectively. Due to significant computational errors, solutions in the set $p_1$
could not be found for $L<75$~nm.

\landscape
\begin{table}[h]
\caption{\bf Values of $ \tq$ computed for $p$-polarized compound SPP waves
in relation to $L$. These values are organized in 5 sets labeled $p_1$ to $p_5$. 
Solutions obtained  for SPP waves guided by either the metal/PMLID interface alone
or  the metal/HID interface alone are also provided, along with the solution for a Tamm
wave guided by HID/PMLID interface when $L=0$~nm.
\label{tab:Psol}
}
\begin{center} 
\begin{tabular}{c|ccccc}
\hline
\cline{2-6}
\rule[-1ex]{0pt}{3.5ex}$L$ (nm) & $p_1$ & $p_2$ & $p_3$ & $p_4$ & $p_5$\\
\hline\hline
\rule[-1ex]{0pt}{3.5ex}$1$ & $\ast$ & $1.8304+ i 1.77\times10^{-4}$ & $-$ & $-$ & $-$\\
\rule[-1ex]{0pt}{3.5ex}$10$ & $\ast$ & $1.8850 + i 3.18\times10^{-3}$ & $-$ & $-$ & $-$\\
\rule[-1ex]{0pt}{3.5ex}$25$ & $\ast$ & $2.0324 + i 1.75\times10^{-2}$ & $-$ & $-$ & $-$\\
\rule[-1ex]{0pt}{3.5ex}$40$ & $\ast$ & $2.2011 + i 4.57\times10^{-2}$ & $-$ & $-$ & $-$\\
\rule[-1ex]{0pt}{3.5ex}$45$ & $\ast$ & $2.2475 + i 5.61\times10^{-2}$ & $1.6708+ i 5.38\times10^{-4}$ & $1.4859+ i 4.93\times10^{-5}$ & $-$\\
\rule[-1ex]{0pt}{3.5ex}$50$ & $\ast$ & $2.2857 + i 6.57\times10^{-2}$ & $1.6711 + i 9.84\times10^{-4}$ & $1.4860+ i 3.76\times10^{-4}$ & $1.2673 + i 3.15\times10^{-5}$\\
\rule[-1ex]{0pt}{3.5ex}$75$ & $2.9595 + i 0.1901$ & $2.3699 + i 9.13\times10^{-2}$ & $1.6715+ i 1.78\times10^{-3}$ & $1.4861+ i 9.68\times10^{-4}$ & $1.2673 + i 1.15\times10^{-4}$\\
\rule[-1ex]{0pt}{3.5ex}$100$ & $2.9458 + i 0.1870$ & $2.3812 + i 9.55\times10^{-2}$ & $1.6716+ i 1.89\times10^{-3}$ & $1.4861 + i 1.05\times10^{-3}$ & $1.2673 + i 1.27\times10^{-4}$\\
\rule[-1ex]{0pt}{3.5ex}$150$ & $2.9446 + i 0.1867$ & $2.3825 + i 9.61\times10^{-2}$ & $1.6716+ i 1.90\times10^{-3}$ & $1.4861 + i 1.07\times10^{-3}$ & $1.2673 + i 1.29\times10^{-4}$\\
\rule[-1ex]{0pt}{3.5ex}$200$ & $2.9446 + i 0.1867$ & $2.3825 + i 9.61\times10^{-2}$ & $1.6716+ i 1.90\times10^{-3}$ & $1.4861 + i 1.07\times10^{-3}$ & $1.2673 + i 1.29\times10^{-4}$\\
\hline\hline
\multicolumn{1}{l}{$\tq_{\small met/PMLID}$} & $2.9446 + i 0.1867$ & {} &  $1.6716 + i 1.90\times10^{-3}$ & $1.4861 + i 1.07\times10^{-3}$ &$1.2673 + i 1.29\times10^{-4}$  \\
\hline
\multicolumn{1}{l}{$\tq_{\small met/HID}$} & {} & $2.3825 + i 9.61\times10^{-2}$ &  {} & {} & {} \\
\hline
\hline
 \multicolumn{1}{c}{$\tq_{\small HID/PMLID}$} & \multicolumn{1}{c}{} & $1.8260$ &\multicolumn{3}{l}{}\\
\hline
\end{tabular}
$^\ast$ No solution was found because of significant numerical error in evaluating $\quadr{{\underline{\underline Y}}^{\tond{p}}(\tq\ko)}$
for ${\rm Re}(\tq)>3.0$.
\end{center}
\end{table}
\endlandscape

\subsubsection{Superset $\Pa$}
The sets $p_1$ to $p_5$ in \tref{tab:Psol} can be grouped into two supersets. Superset $\Pa$ comprises the sets
$p_1$, $p_3$, $p_4$, and $p_5$. As $L$ increases, the wavenumber of a $p$-polarized compound SPP wave in any of the 
four sets in $\Pa$ tends
towards the wavenumber of a $p$-polarized  SPP wave that can be guided by the metal/PMLID interface all
by itself. As $L$ decreases, both the phase speed $\vph=c\ped0/{\rm Re}(\tq)$
and the propagation distance $\propdist=1/\ko{\rm Im}(\tq)$ decrease in set $p_1$;
both $\vph$ and $\propdist$ increase in the sets $p_3$ and $p_4$; while $v_{ph}$ is practically constant and  $\propdist$ increases in the set $p_5$.

 \begin{figure}
 \begin{center}
 \begin{tabular}{c}
(a) \includegraphics[width=0.35\linewidth]{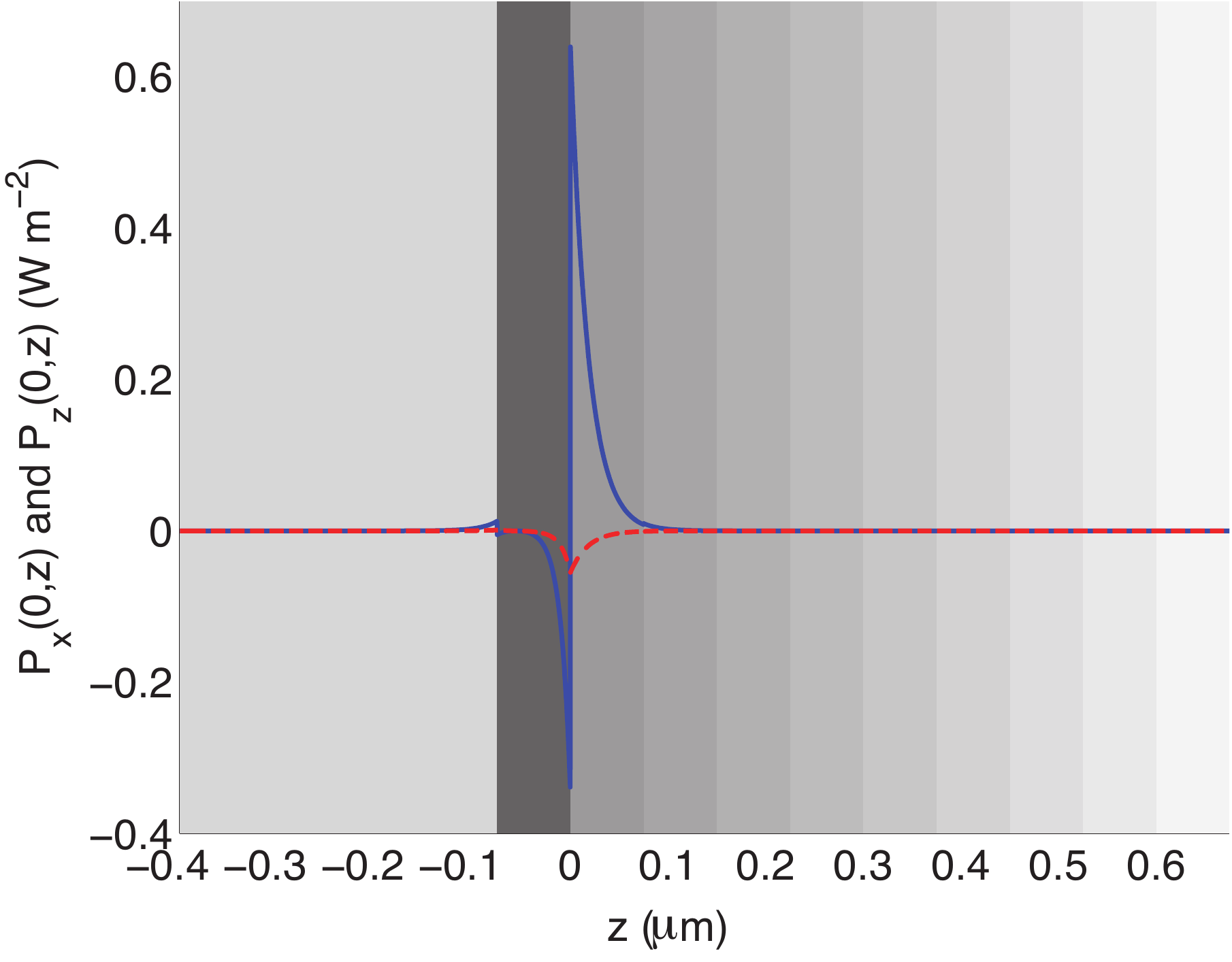}\hspace{10pt}
(b)  \includegraphics[width=0.35\linewidth]{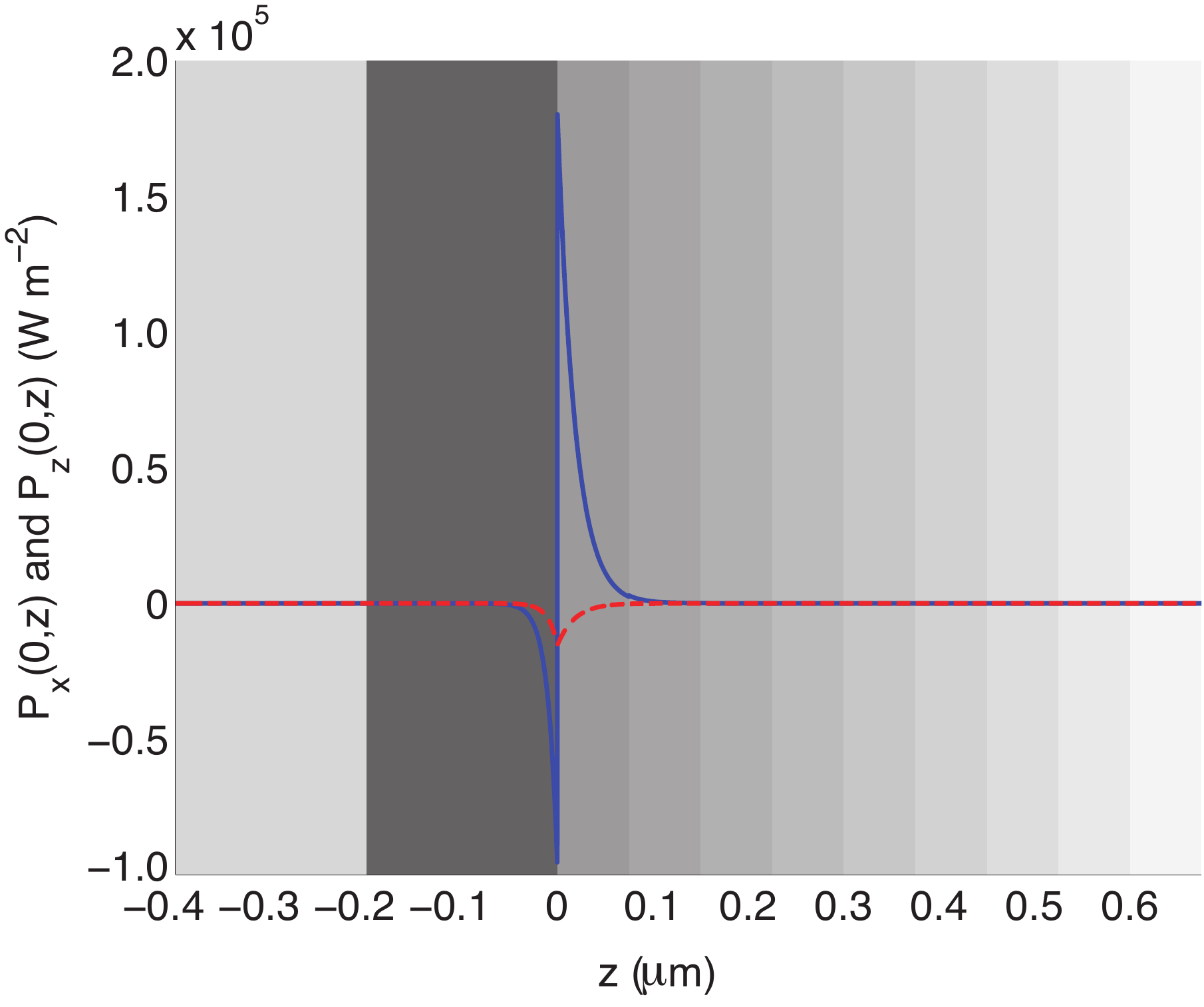}\\
(c) \includegraphics[width=0.35\linewidth]{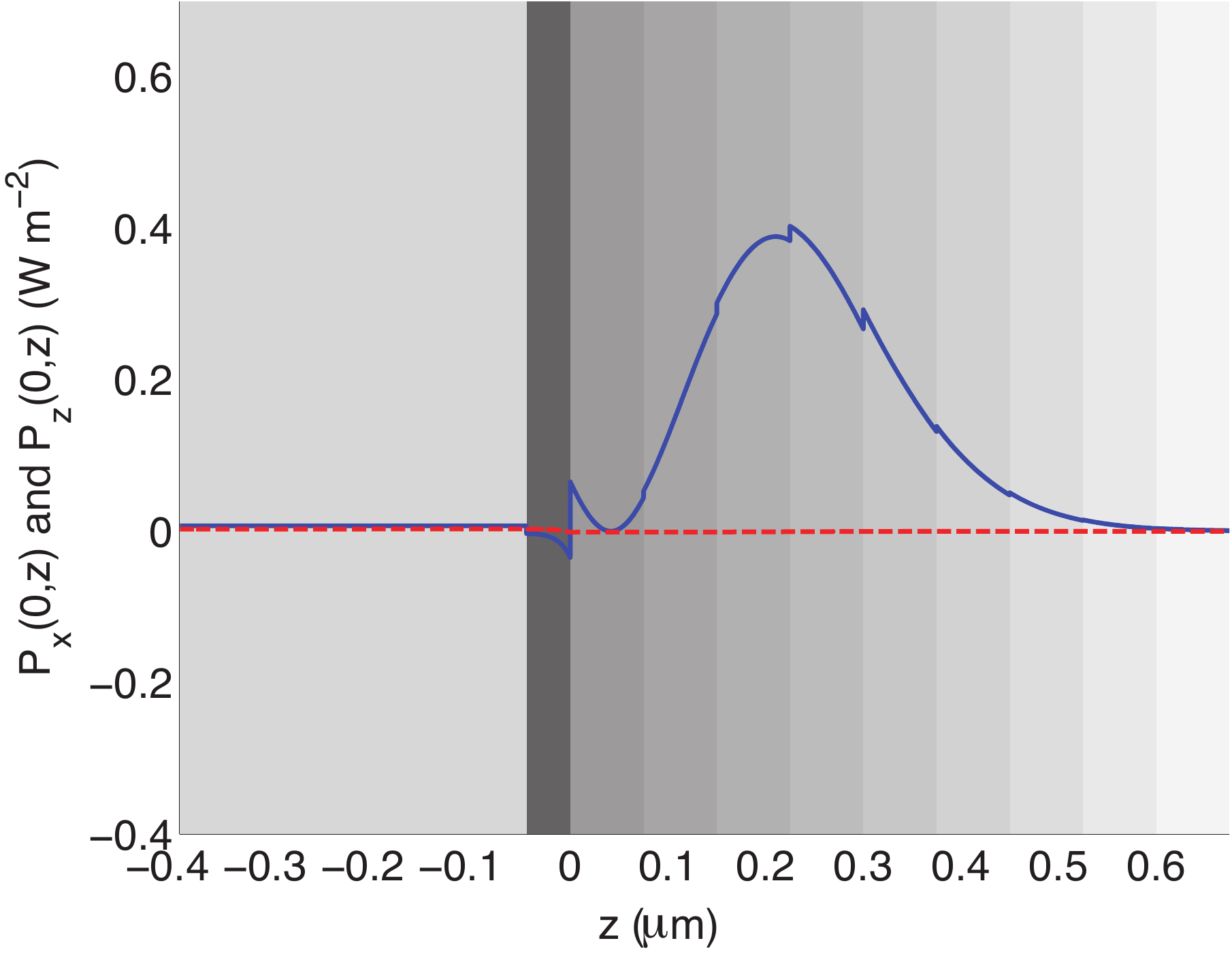}\hspace{10pt}
(d) \includegraphics[width=0.35\linewidth]{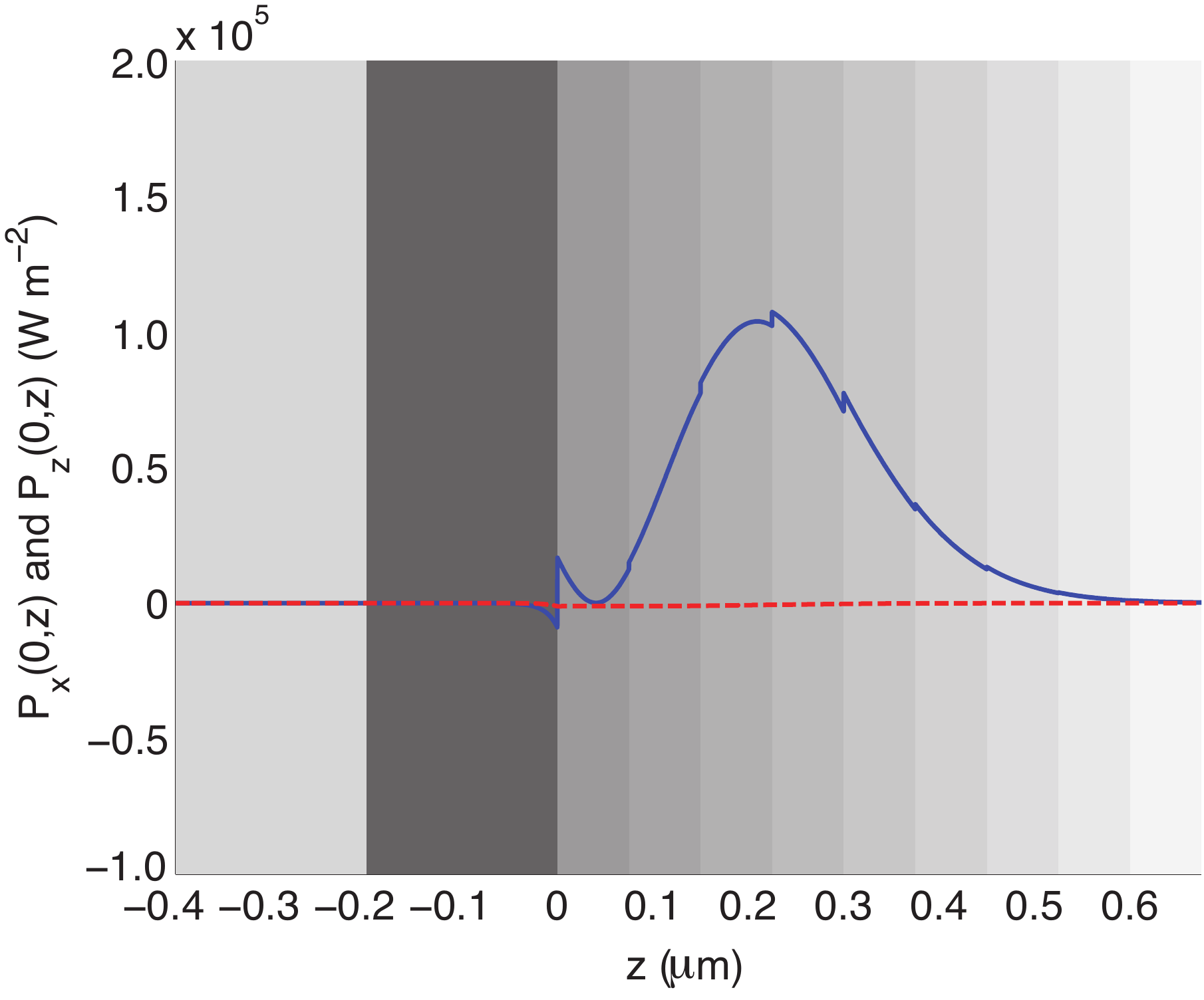}\\
(e) \includegraphics[width=0.35\linewidth]{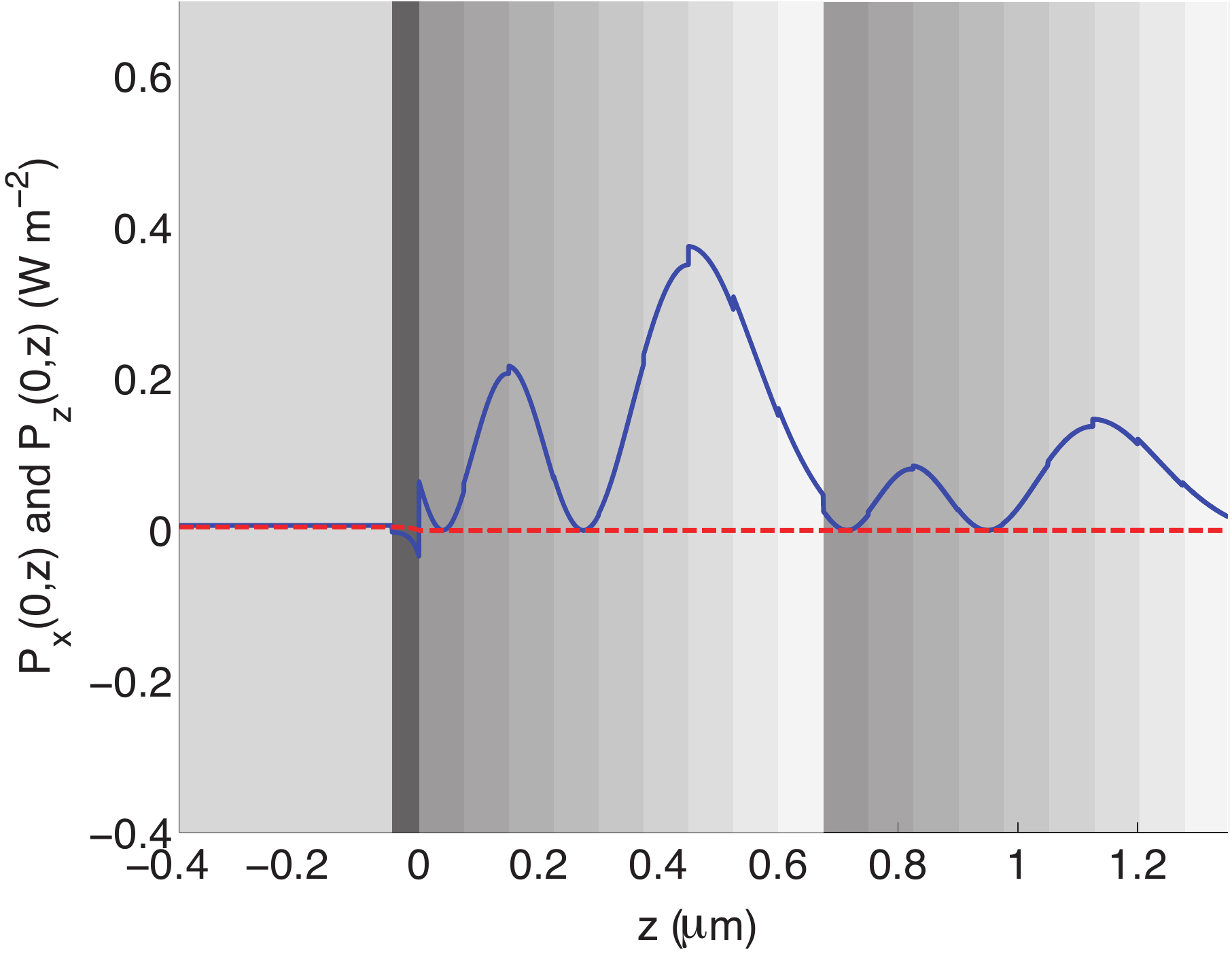}\hspace{10pt}
(f) \includegraphics[width=0.35\linewidth]{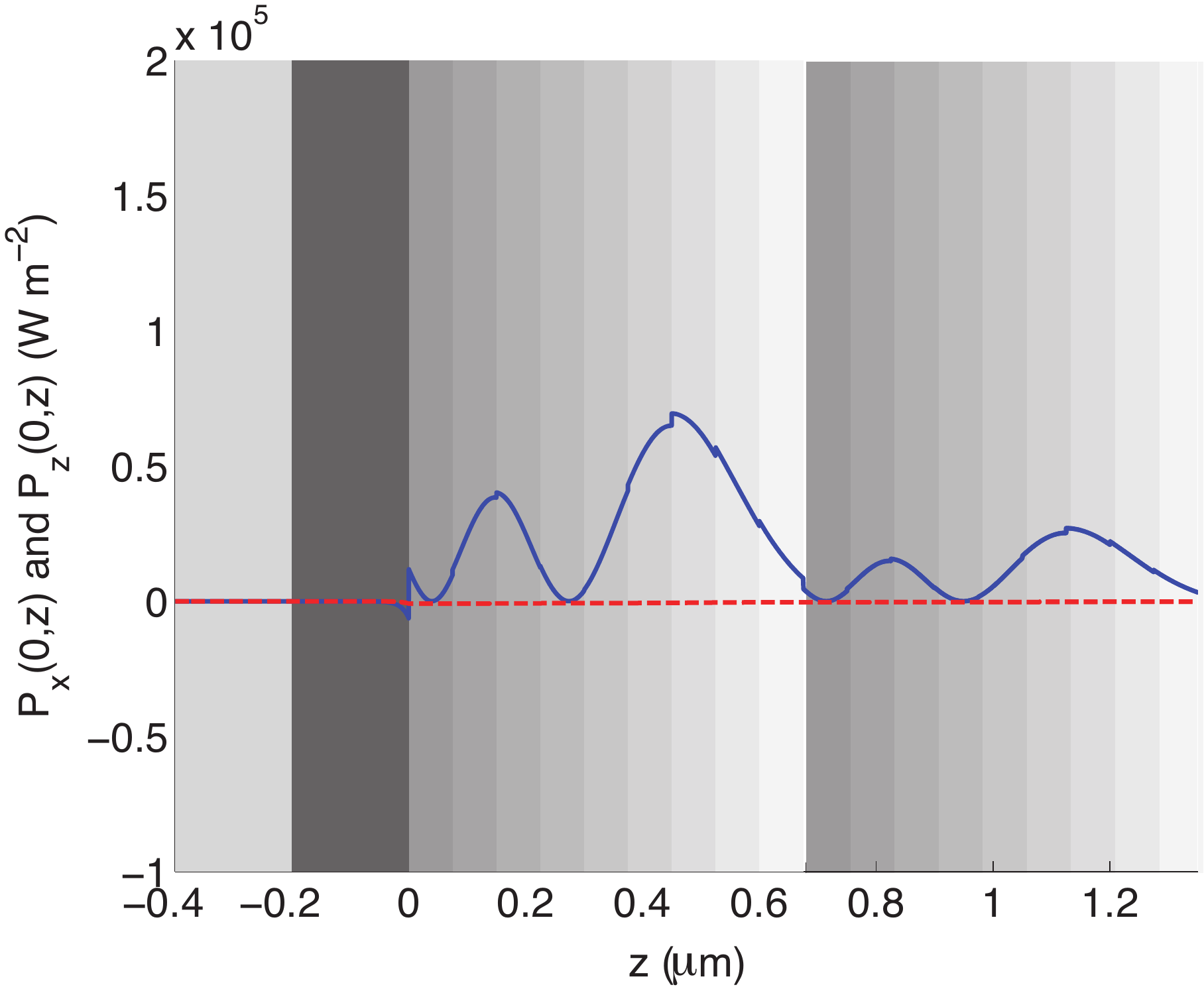}\\
(g) \includegraphics[width=0.35\linewidth]{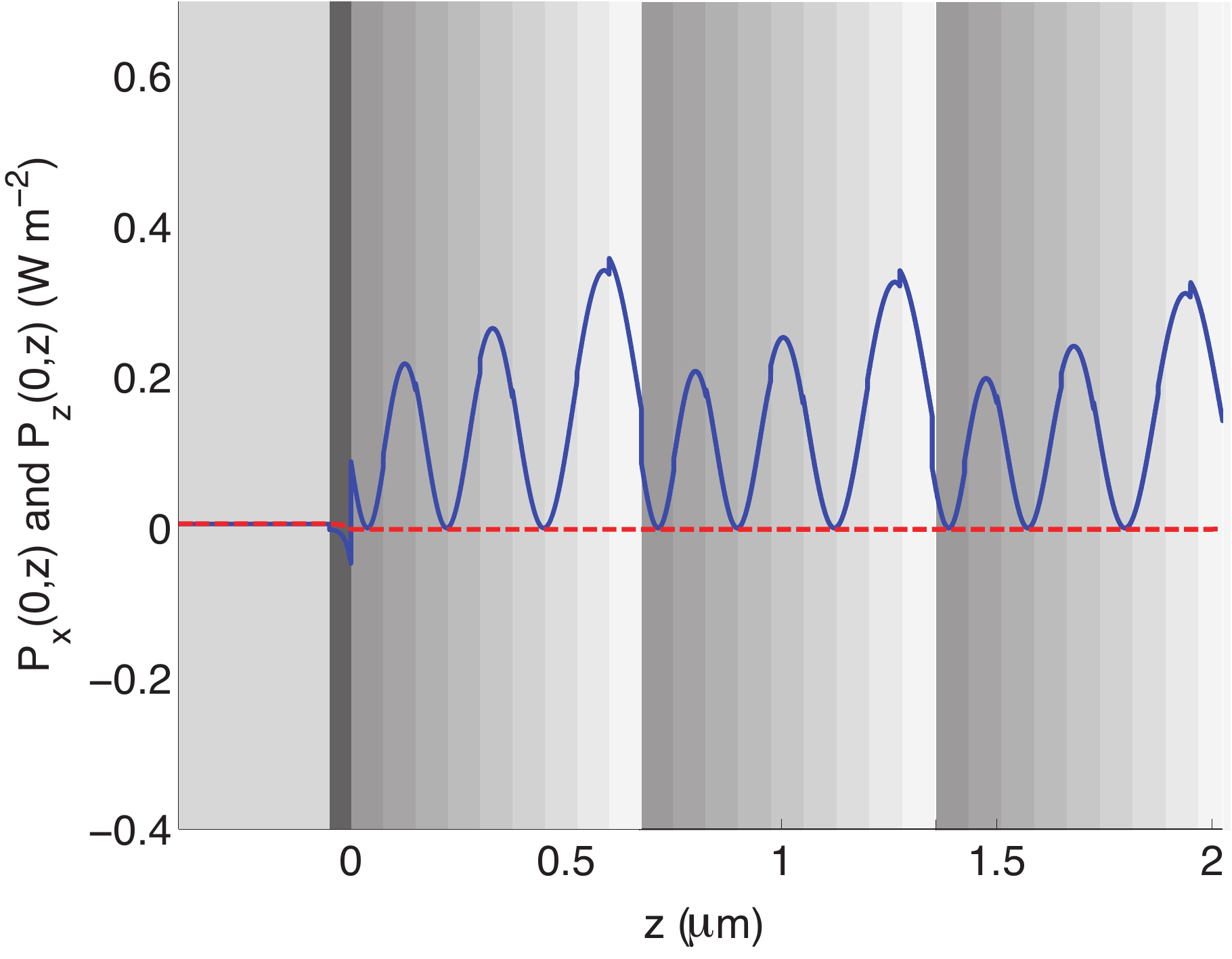}\hspace{10pt}
(h) \includegraphics[width=0.35\linewidth]{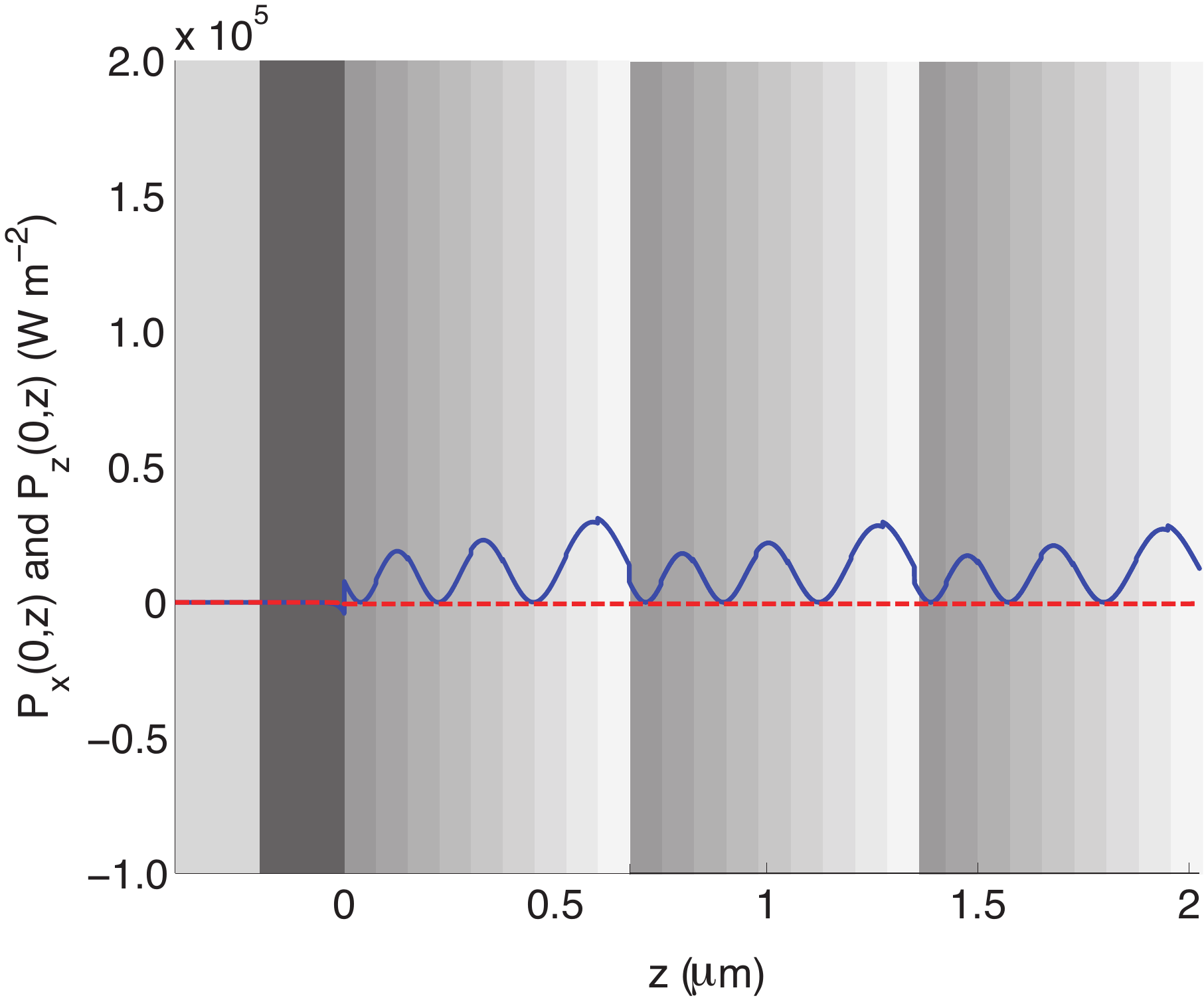}
   \end{tabular}
 \end{center}
\caption{
Spatial variations of $P_x(0,z)$ (blue slide lines) and $P_z(0,z)$ (red dashed lines) with respect to $z$
for the compound $p$-polarized SPP waves in Superset $\Pa$.
(a) $L=75$~nm or (b) $L=200$~nm in set $p_1\subset\Pa$;
(c) $L=45$~nm or (d) $L=200$~nm in set $p_3\subset\Pa$;
(e) $L=45$~nm or (f) $L=200$~nm in set $p_4\subset\Pa$; and
(g) $L=50$~nm or (h) $L=200$~nm in set $p_5\subset\Pa$.
 \label{fig:SupersetPa}}
 \end{figure} 

Figure~\ref{fig:SupersetPa} shows the spatial variations of the Cartesian components $P_x(0,z)$ and $P_z(0,z)$ of the
time-averaged Poynting vector  in the plane $x=0$ for representative compound SPP waves belonging to the four sets in $\Pa$. Whereas $P_x(0,z)$ can be discontinuous across bilateral interfaces due to the discontinuity of $e_z$ defined in \eref{eq:Ez} across those
interfaces, $P_z(0,z)$ must be a continuous function of $z$. These characteristic features are evident in
Fig.~\ref{fig:SupersetPa}, which also demonstrates that the compound SPP waves are bound predominantly to
the metal/PMLID interface. The power density is 
confined mostly  to the PMLID material: specifically, the first unit cell for compound SPP waves in sets $p_1$ and
$p_3$, and more than one unit cell for compound SPP waves in sets $p_4$ and
$p_5$.  The affinity to the SPP wave guided by the metal/PMLID interface by itself is particularly pronounced  when the ratio $L/\delta_m$ is large, indicative of weak coupling  between the two metal/dielectric interfaces. 

As $L$ decreases, the compound SPP wave belonging to the set $p_1$  becomes also bound to the metal/HID interface,
thereby indicating a significant coupling between the two metal/dielectric interfaces. However, compound SPP waves belonging
to $p_3$, $p_4$, and $p_5$ cease to exist when $L/\delta_m \lesssim 1.4$ rather than divert significant energy from the PMLID
material to the HID material.

\subsubsection{Superset $\Pb$}
Superset $\Pb$ comprises only the set $p_2$. As $L$ increases, the value of $\tq$ approaches the value of $\tq$ for the sole SPP wave guided by the metal/HID interface all by itself. However, as $L$ decreases,  the value of $\tq$ approaches the value of $\tq$ for a $p$-polarized SPP wave guided by the HID/PMLID interface (i.e., $L=0$). 
The set $p_2$ thus exemplifies a transition, and therefore the inherent commonality, between two surface-wave phenomenons that otherwise appear to be completely different from each other, and adds to the previously found  unity of Fano waves and Tamm waves \cite{FML}. Furthermore, both $\vph$ and $\propdist$ increase in the set $p_2$ as $L$ decreases.

 \begin{figure}
 \begin{center}
 \begin{tabular}{c}
(a) \includegraphics[width=0.35\linewidth]{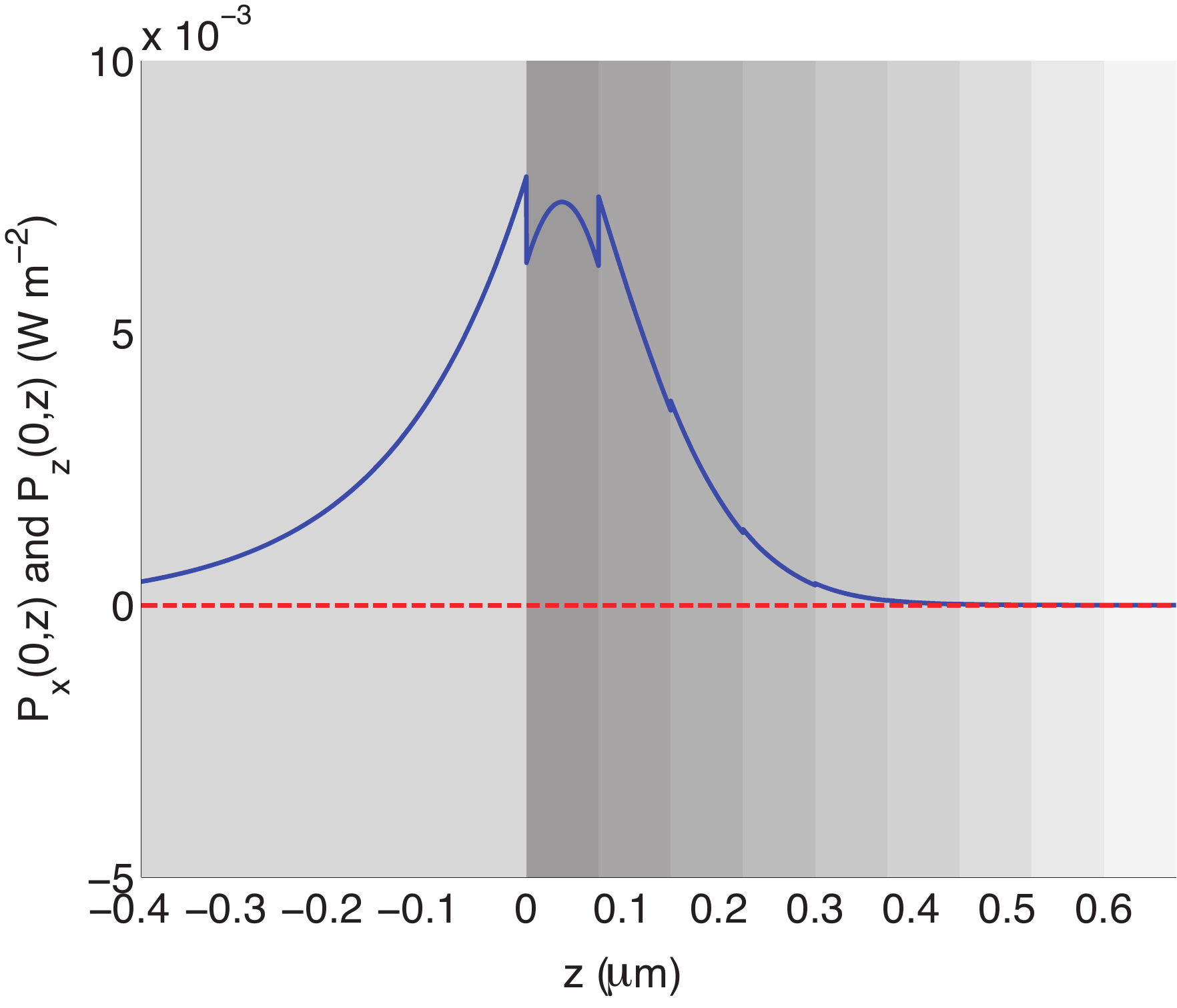}\hspace{10pt}
(b)  \includegraphics[width=0.35\linewidth]{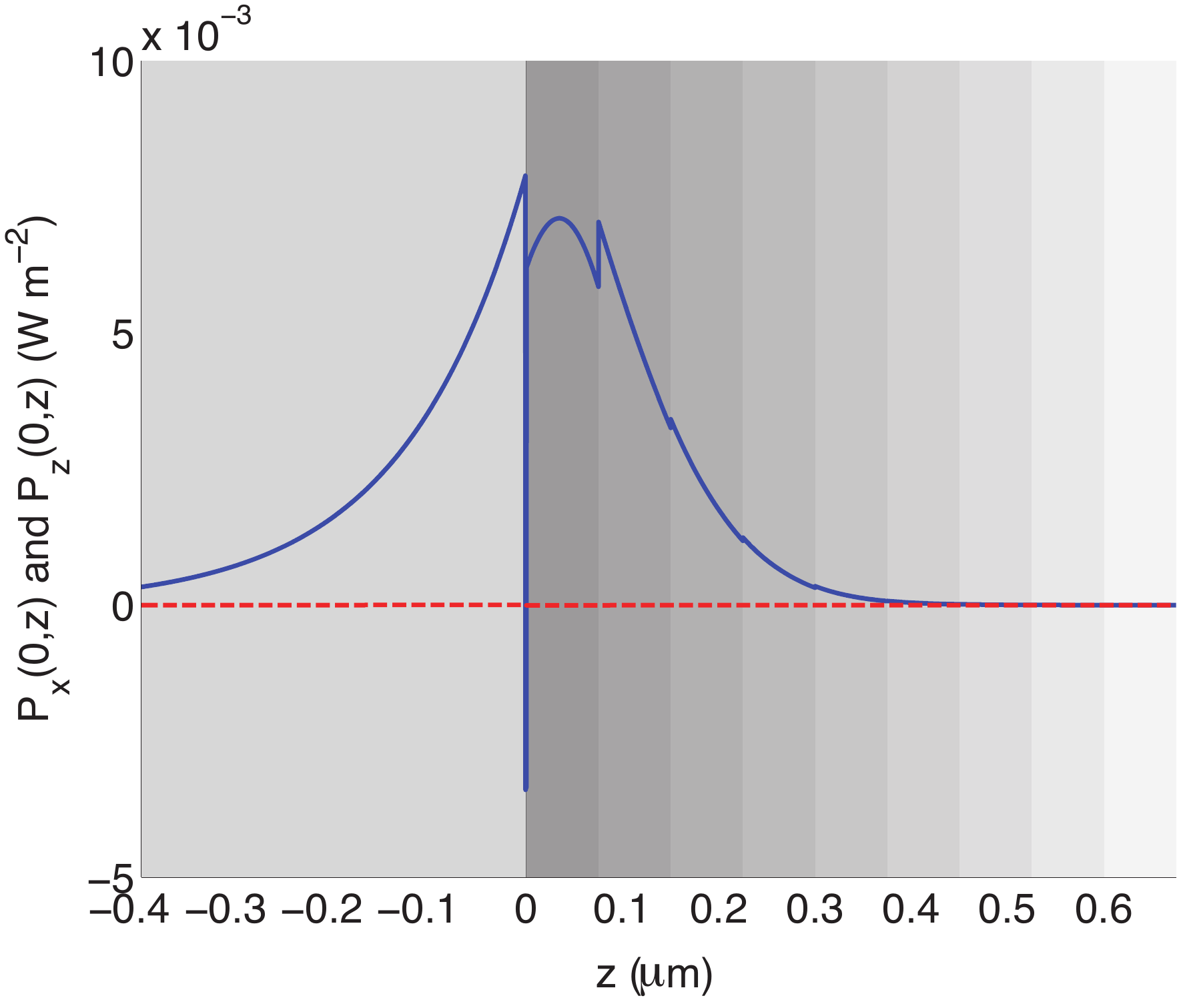}\\
(c) \includegraphics[width=0.35\linewidth]{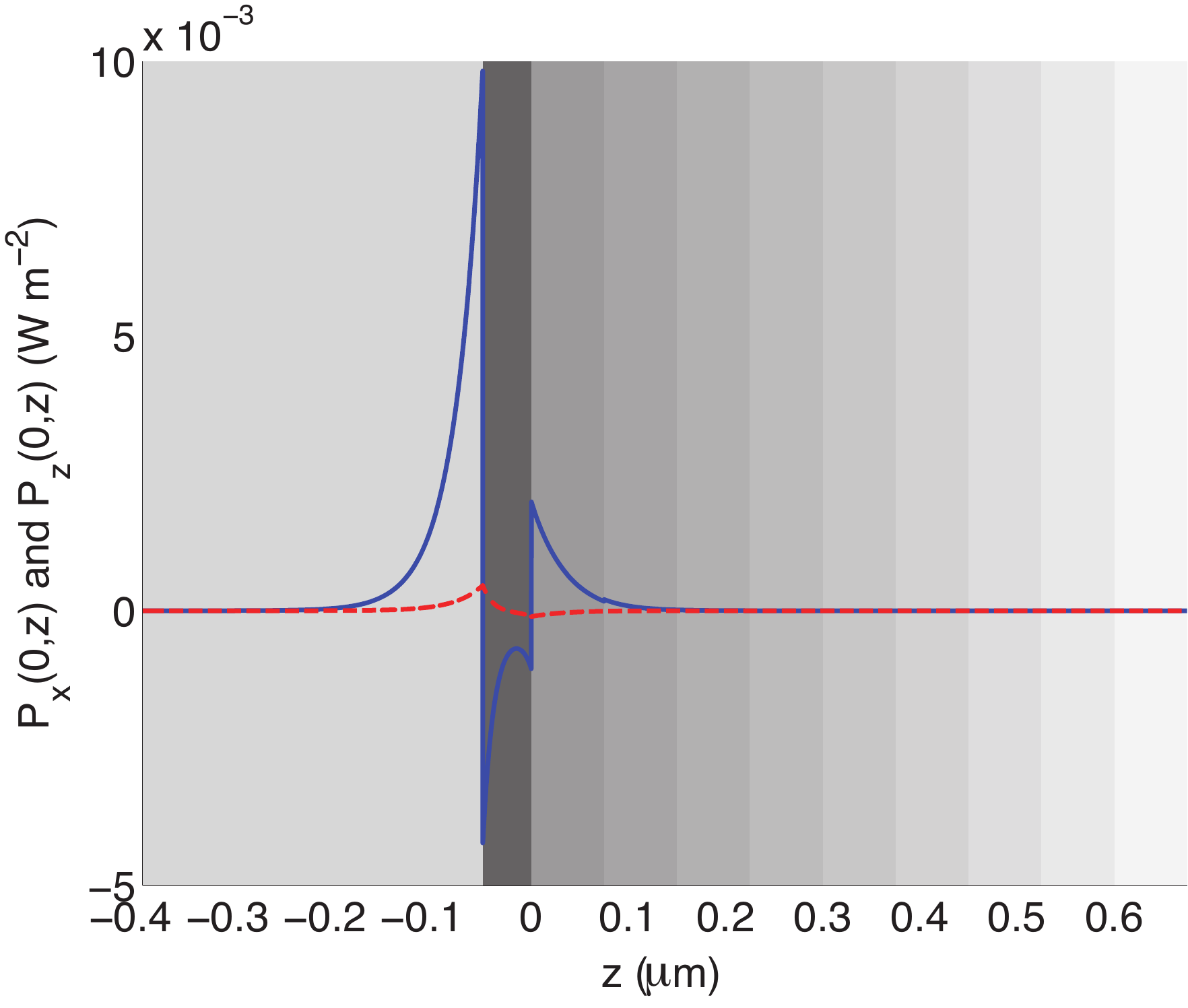}\hspace{10pt}
(d) \includegraphics[width=0.35\linewidth]{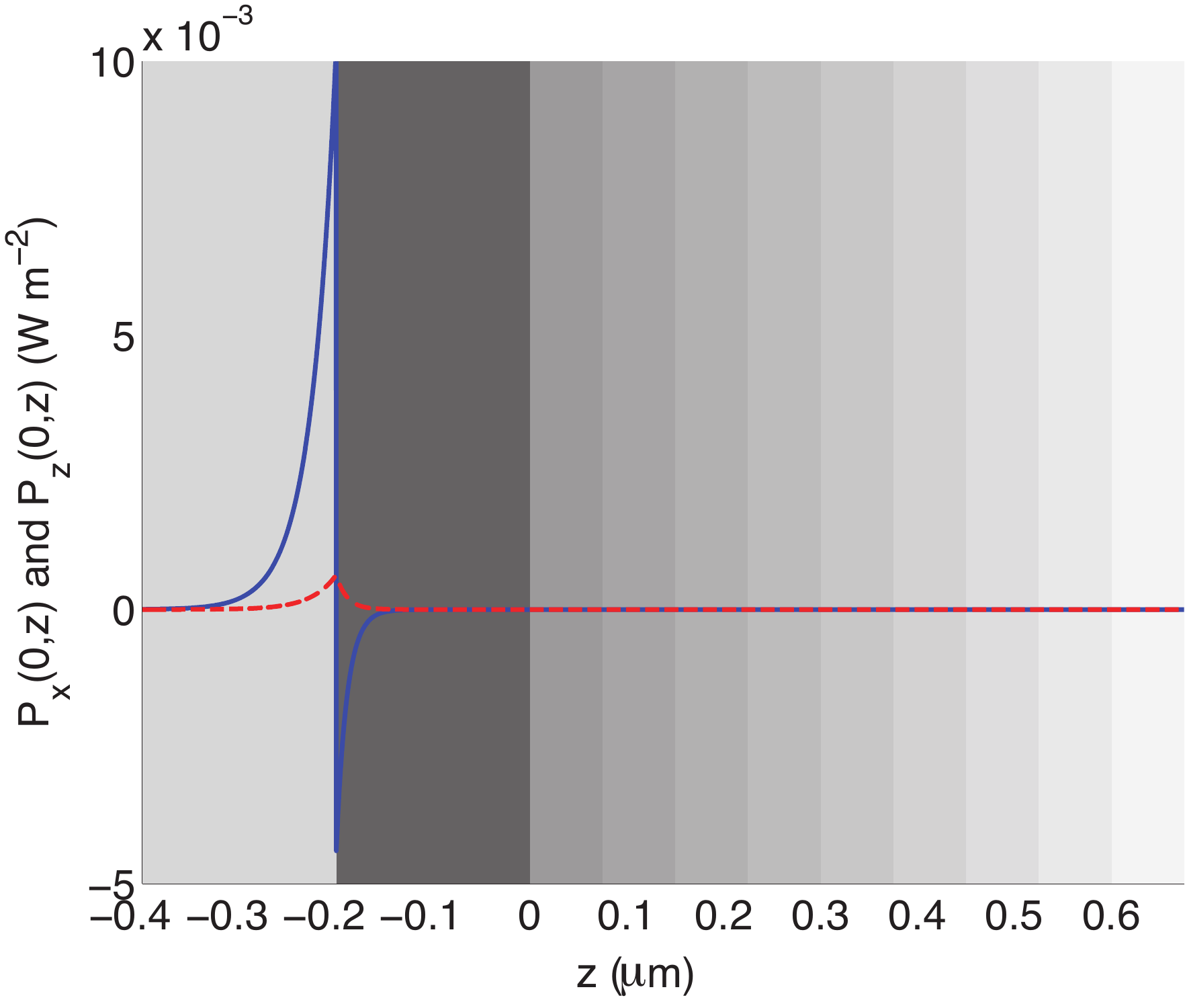}
   \end{tabular}
 \end{center}
\caption{
Spatial variations of $P_x(0,z)$ (blue slide lines) and $P_z(0,z)$ (red dashed lines) with respect to $z$
for the compound $p$-polarized SPP waves in Superset $\Pb$.
(a) $L=0$~nm,  (b) $L=1$~nm, (c) $L=50$~nm, or (d) $L=200$~nm in set $p_2$.
 \label{fig:SupersetPb}}
 \end{figure} 
 
Figure~\ref{fig:SupersetPb} shows the spatial variations of   $P_x(0,z)$ and $P_z(0,z)$ for representative compound SPP waves belonging to  $p_2\subseteq\Pb$. Again, while $P_x(0,z)$ is discontinuous across bilateral interfaces, $P_z(0,z)$ is a continuous function of $z$. When the ratio $L/\delta_m$ is large, the compound SPP wave is predominantly bound to the metal/HID interface. As $L/\delta_m$ decreases, the compound SPP wave also gets bound  to the metal/PMLID interface and the  energy begins to be transferred from the HID material to the PMLID material. Of course, for $L=0$ the surface wave is bound to the HID/PMLID interface and is transformed into a Tamm wave \cite{YYH,YYC,AkhBook}.

\subsection{$s$-polarized compound SPP waves}\label{sSPPres}
The normalized wavenumbers $ \tilde{q} $ calculated for $s$-polarized compound SPP waves are reported in \tref{tab:Ssol} for different  thicknesses $L \in\left[30,200 \right]$~nm of the metal layer.  The table also lists the values of the three $s$-polarized SPP waves guided by the metal/PMLID interface by itself and the value of the Tamm wave guided by the 
HID/PMLID interface. No $s$-polarized SPP wave can be guided by the metal/HID interface by itself \cite{Pitarke,AkhBook}.

According to the table, as many as three different  compound $s$-polarized SPP waves can be guided by the 
HID/metal/PMLID structure for  $L \in\left[30,200 \right]$~nm.  These waves are organized in three sets labeled $ s_1 $ to $ s_3 $. Compound s-polarized SPP waves do not exist if the thickness $ L $ of the metal layer is  
$ \lesssim $  $ 29.3 $, $ 39.3 $, and $ 43.8 $~nm for the sets $ s_1 $, $ s_2 $, and $ s_3 $, respectively. Therefore it is not surprising that, unlike for compound $p$-polarized SPP waves,  the normalized wavenumber $\tq$ in none of the three sets of solutions approaches the value of  $\tq$ for  the Tamm wave as $ L $ decreases. 

Because the metal/HID interface by itself  cannot guide $s$-polarized SPP waves,   the wavenumber of  the compound SPP waves in any of the three sets tends towards the wavenumber of an $s$-polarized  SPP wave guided by the metal/PMLID interface  by itself as $L$ increases.
Finally,  as $L $ decreases, the phase speed  $\vph$ decreases but the propagation distance $\propdist$ increases in all three sets $s_1$ to $s_3$.  

\begin{figure}
  \begin{center}
  \begin{tabular}{c}
  (a) \includegraphics[width=0.35\linewidth]{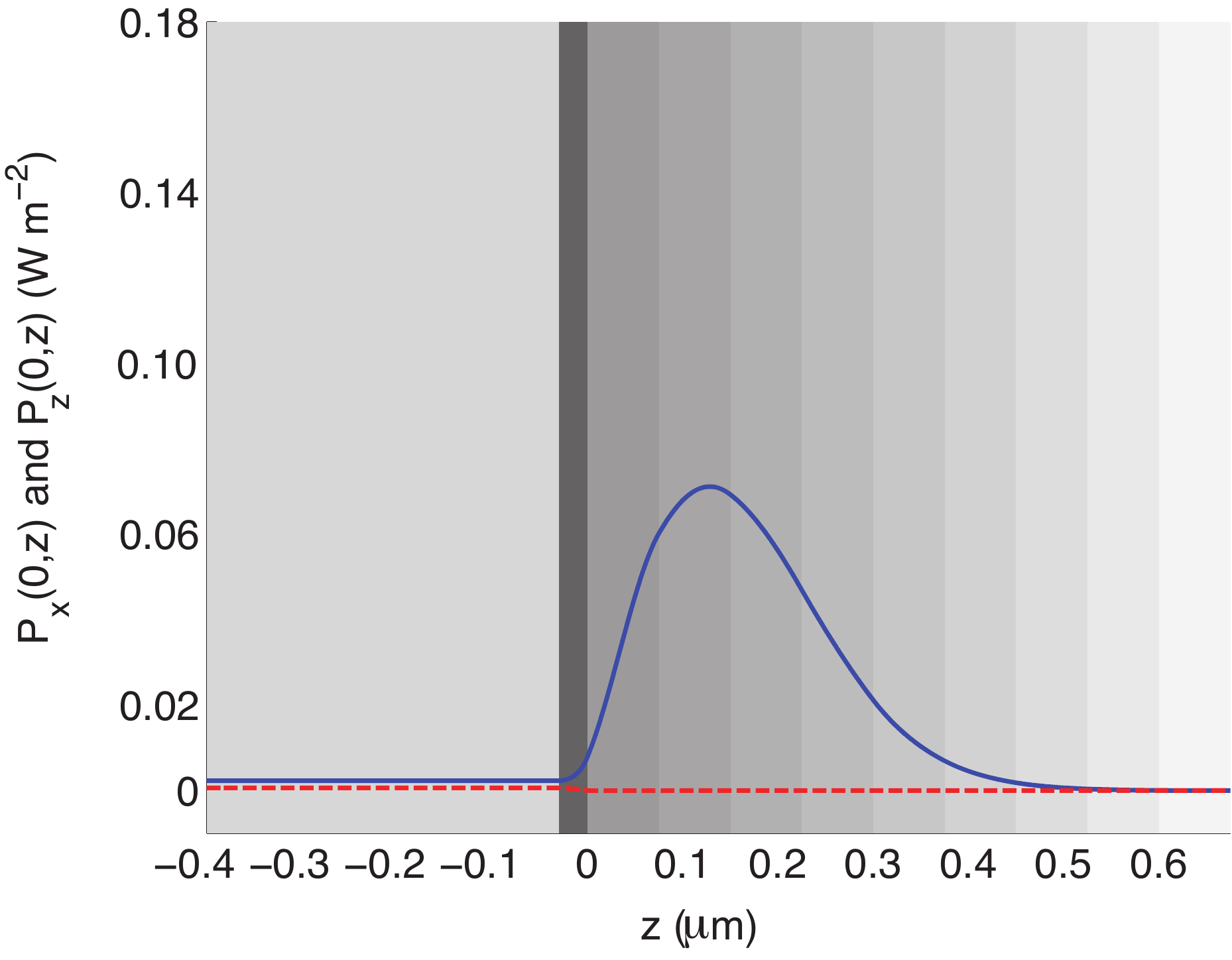}\hspace{10pt}
  (b)  \includegraphics[width=0.35\linewidth]{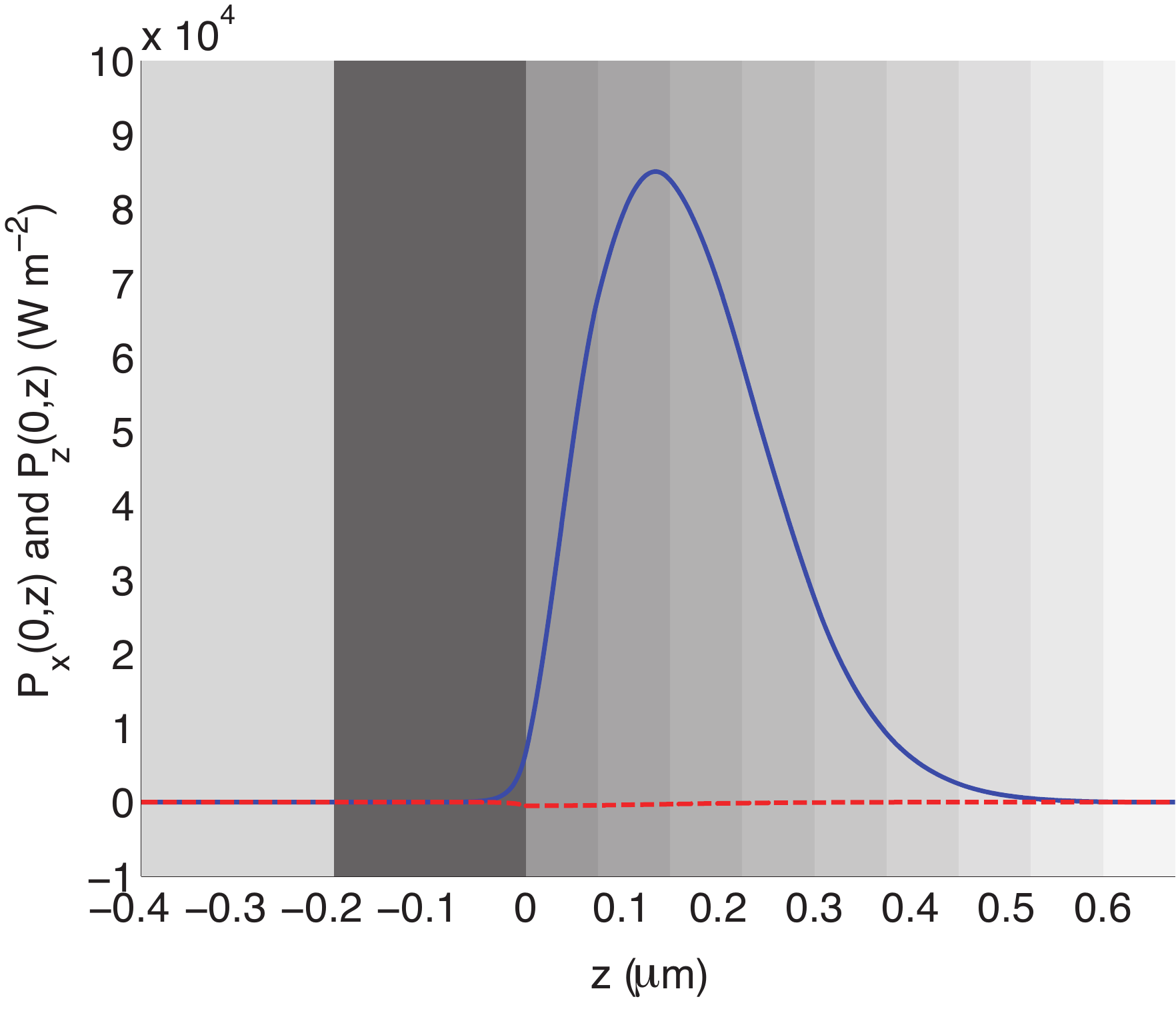}\\
  (c) \includegraphics[width=0.35\linewidth]{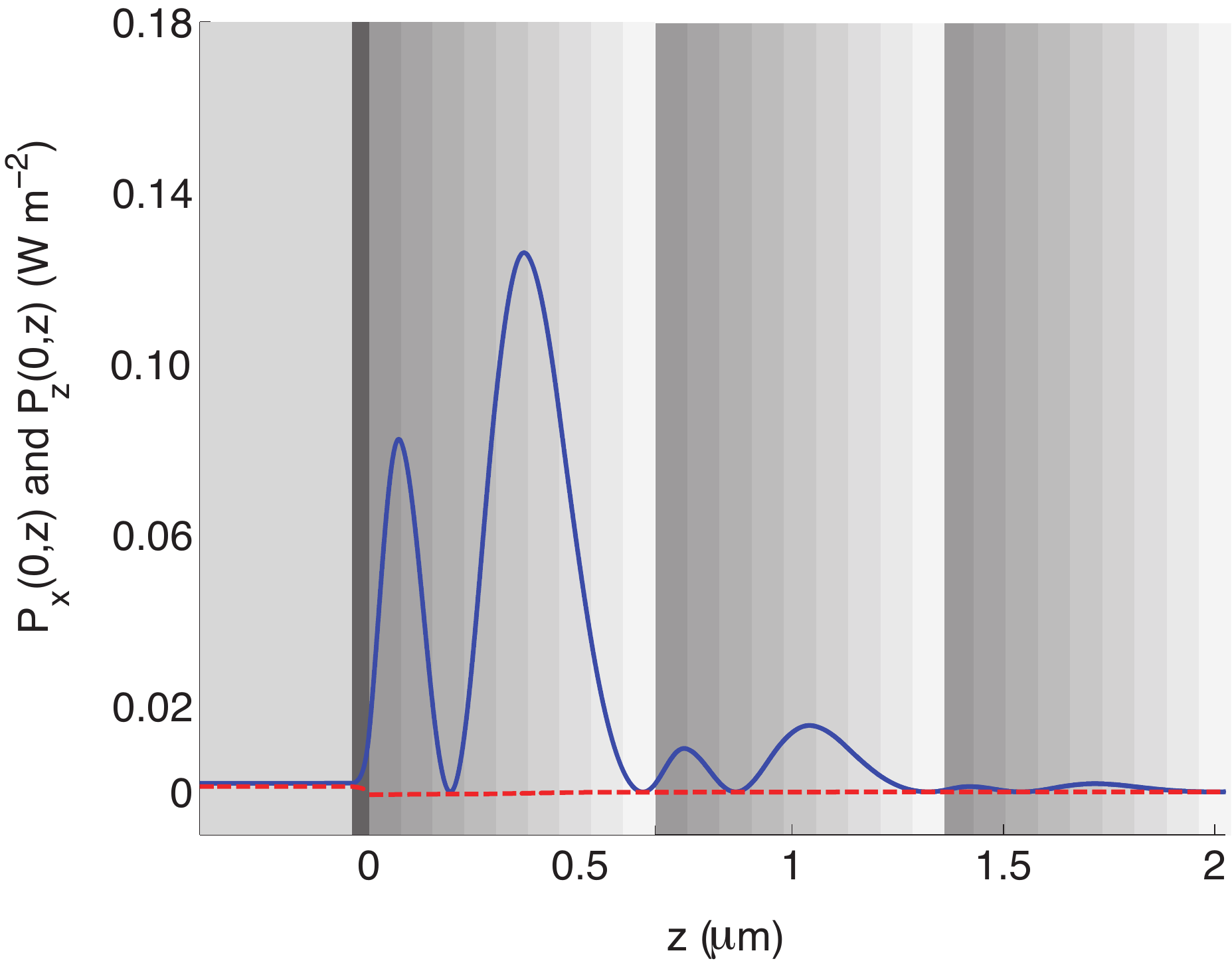}\hspace{10pt}
  (d) \includegraphics[width=0.35\linewidth]{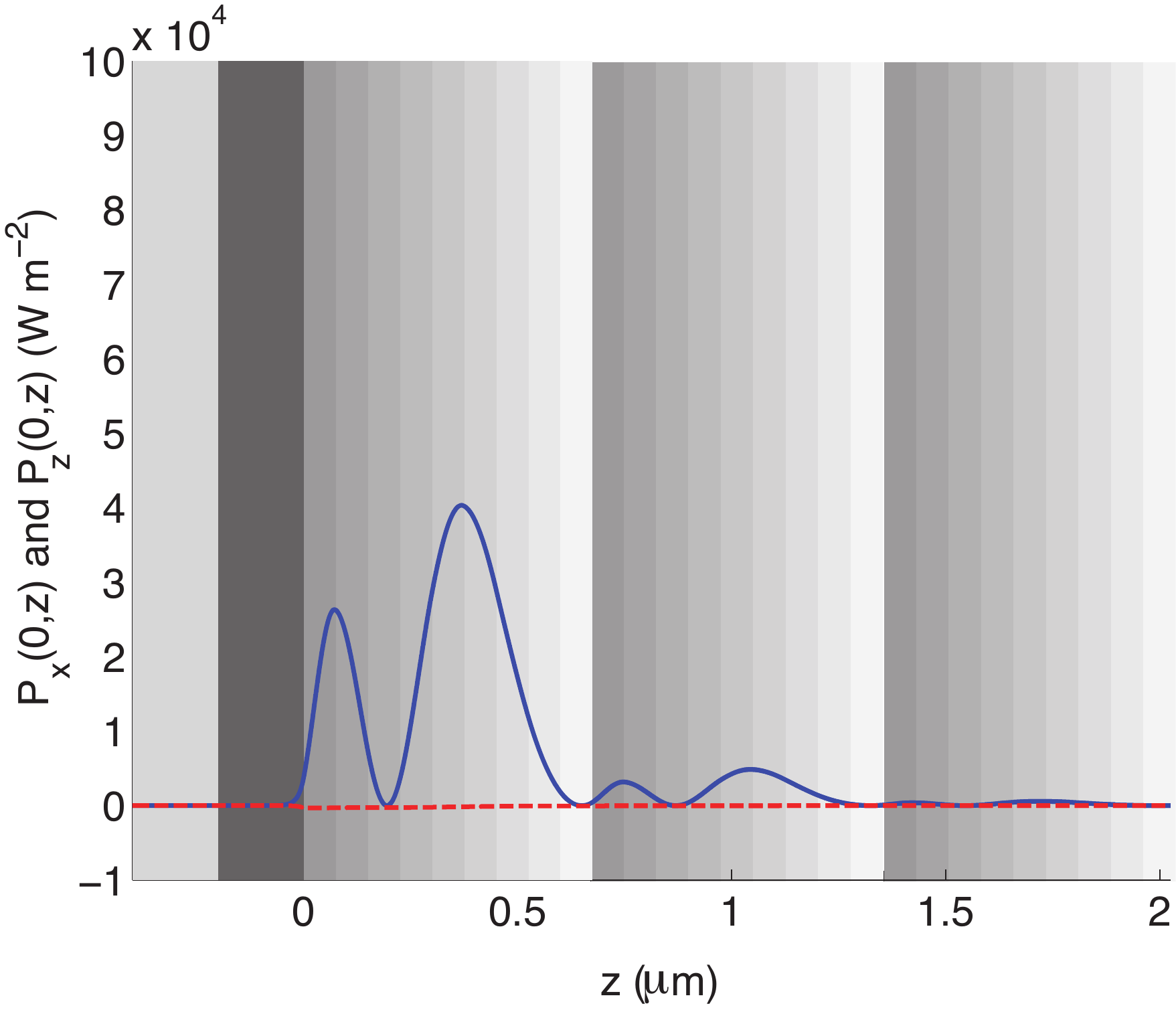}\\
  (e) \includegraphics[width=0.35\linewidth]{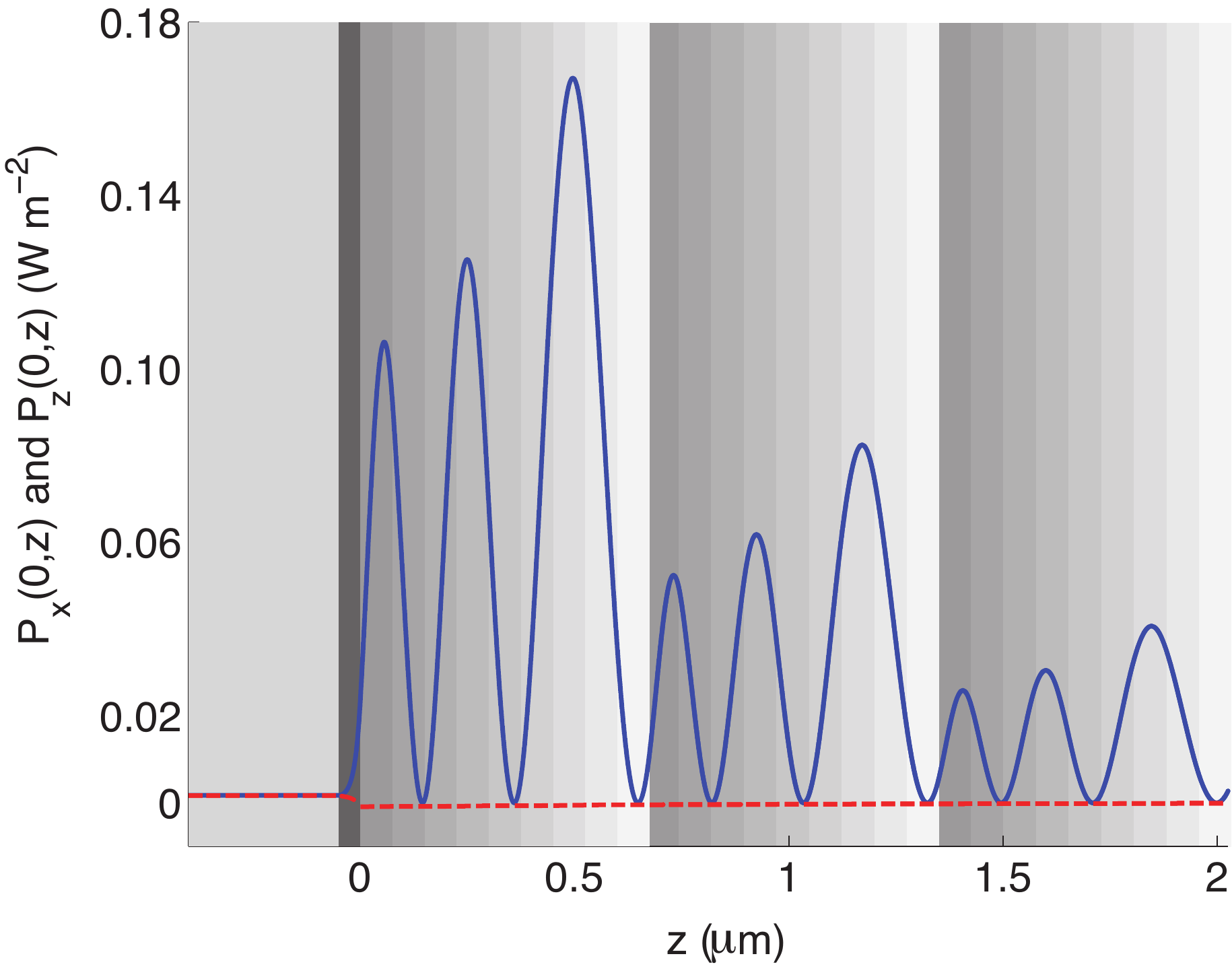}\hspace{10pt}
  (f) \includegraphics[width=0.35\linewidth]{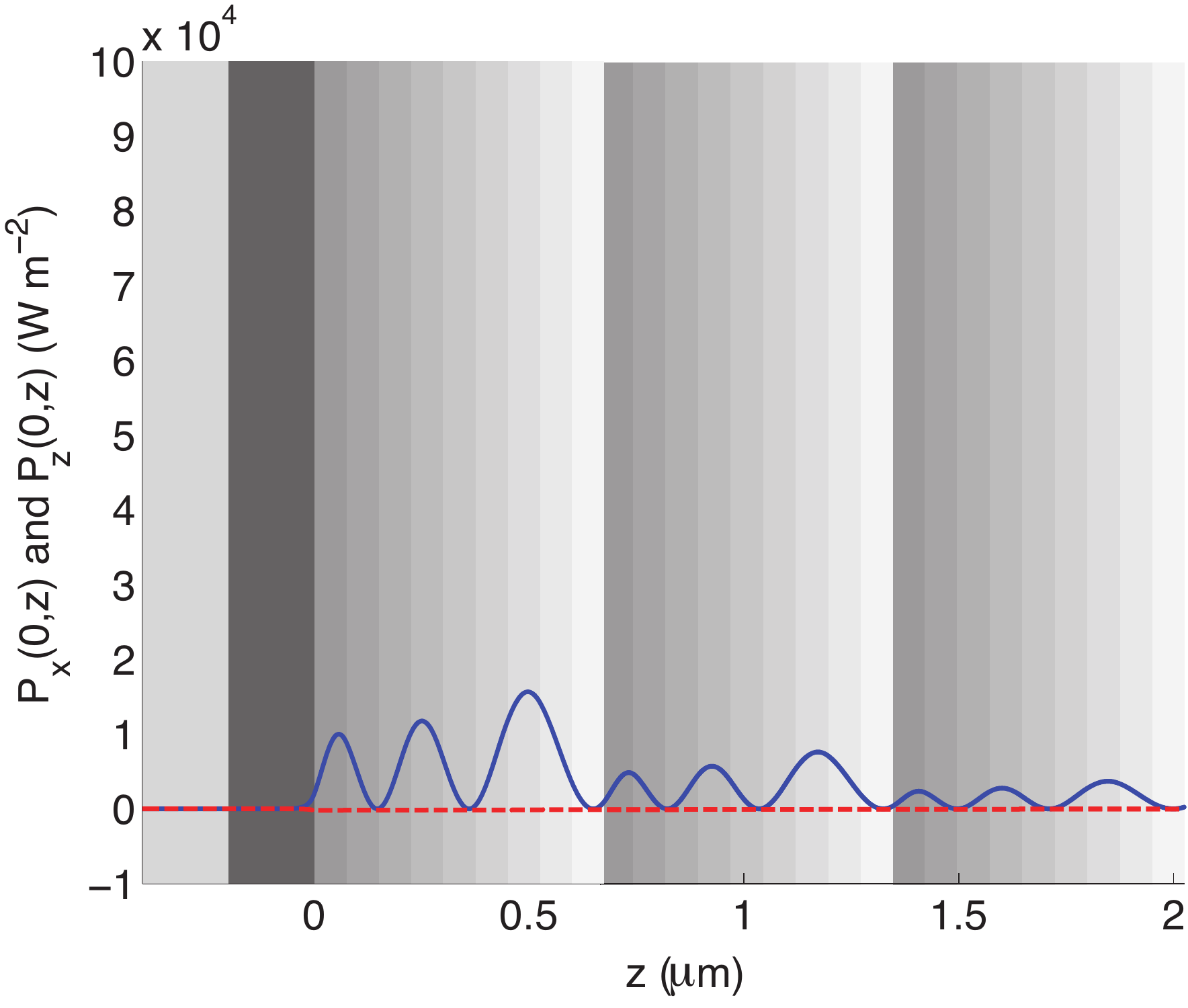}\\
   \end{tabular}
\caption{
Spatial variations of $P_x(0,z)$ (blue solid lines) and $P_z(0,z)$ (red dashed lines) with respect to $z$
for the compound $s$-polarized SPP waves.
(a) $L=30$~nm or (b) $L=200$~nm in set $s_1$;
(c) $L=40$~nm or (d) $L=200$~nm in set $s_2$; and
(e) $L=50$~nm or (f) $L=200$~nm in set $s_3$.
 \label{fig:SupersetS}}
  \end{center}
  \end{figure}

\fref{fig:SupersetS} shows the spatial variation  of the Cartesian components  $ P_x\left(0,z \right) $ and $ P_z\left(0,z \right) $ of the time-averaged Poynting vector in the plane $ x=0 $ for representative compound SPP waves belonging to the three sets of solutions. Unlike for  compound $p$-polarized SPP waves, both components are 
continuous functions of $z$ due to the continuity across interfaces of $e_y(z)$, $h_x(z)$ and, in particular,
$h_z (z)$ by virtue of  \eref{eq:hz}. The
figure shows that the compound $s$-polarized SPP waves are bound almost totally to the metal/PMLID interface resulting in a negligible coupling between the two metal/dielectric interfaces. As a consequence,  the power density is confined to the PMLID material: specifically, the first unit cell for compound SPP waves in the set $s_1$, and more than one unit cell for compound SPP waves in the sets $s_2$ and $s_3$.

\begin{table}[h]
\caption{\bf Values of $ \tq$ computed for $s$-polarized compound SPP waves in relation to $L$. These values are organized in 3 sets labeled $s_1$ to $s_3$. 
Solutions obtained  for SPP waves guided by the metal/PMLID interface alone are also provided, along with the solution for a Tamm
wave guided by HID/PMLID interface when $L=0$~nm. 
\label{tab:Ssol}
}
\begin{center} 
\begin{tabular}{c|ccc}
\hline
\cline{2-4}
\rule[-1ex]{0pt}{3.5ex}$L$ (nm) & $s_1$ & $s_2$ & $s_3$\\
\hline\hline
\rule[-1ex]{0pt}{3.5ex} $30$ & $1.7375 + i 8.13\times10^{-5}$ & $-$ & $-$\\
\rule[-1ex]{0pt}{3.5ex} $40$ & $1.7347 + i 6.85\times10^{-4}$ & $1.5403 + i 6.44\times10^{-4}$ & $-$\\
\rule[-1ex]{0pt}{3.5ex} $50$ & $1.7335 + i 8.91\times10^{-4}$ & $1.5392 + i 6.02\times10^{-4}$ & $1.3066 + i 3.87\times10^{-4}$\\
\rule[-1ex]{0pt}{3.5ex} $75$ & $1.7327 + i 9.93\times10^{-4}$ & $1.5384 + i 9.37\times10^{-4}$ & $1.3059 + i 8.95\times10^{-4}$\\
\rule[-1ex]{0pt}{3.5ex} $100$ & $1.7326 + i 9.98\times10^{-4}$ & $1.5383 + i 9.74\times10^{-4}$ & $1.3057 + i 9.62\times10^{-4}$\\
\rule[-1ex]{0pt}{3.5ex} $150$ & $1.7326+ i 9.98\times10^{-4}$ & $1.5383 + i 9.78\times10^{-4}$ & $1.3057 + i 9.71\times10^{-4}$\\
\rule[-1ex]{0pt}{3.5ex} $200$ & $1.7326 + i 9.98\times10^{-4}$ & $1.5383 + i 9.78\times10^{-4}$ & $1.3057 + i 9.71\times10^{-4}$\\
\hline\hline
$\tq_{\small met/PMLID}$ & $1.7326 + i 9.98\times 10^{-4}$ & $1.5383 + i 9.78\times10^{-4}$ & $1.3057 + i 9.71\times 10^{-4}$\\
\hline\hline
\multicolumn{1}{l}{$\tq_{\small HID/PMLID}$} & \multicolumn{3}{c}{$1.8375$}\\
\hline
\end{tabular}
\end{center}
\end{table}

\subsection{Effect of the relative permittivity of the HID material}\label{variableHID}
The data provided in Secs.~\ref{pSPPres} and \ref{sSPPres}  demonstrate that when the thickness $L$ of the  metal layer  is small enough  the two dielectric/metal interfaces  are significantly coupled each other and  compound SPP waves propagate bound to both   interfaces. 
Moreover,   only one compound SPP
wave (in set $p_2$)  exists for $L$ smaller than the skin depth $\delta_m$ in  the metal. But this paucity of compound SPP waves when $L\lesssim\delta_m$ is not a universal feature, as became clear when we
fixed $L = 25$~nm but varied
 the relative permittivity $\eps_d$ of the HID material from $1$ to $4$. 
 
\begin{figure}
 \begin{center}
 \begin{tabular}{c}
\includegraphics[width=0.5\linewidth]{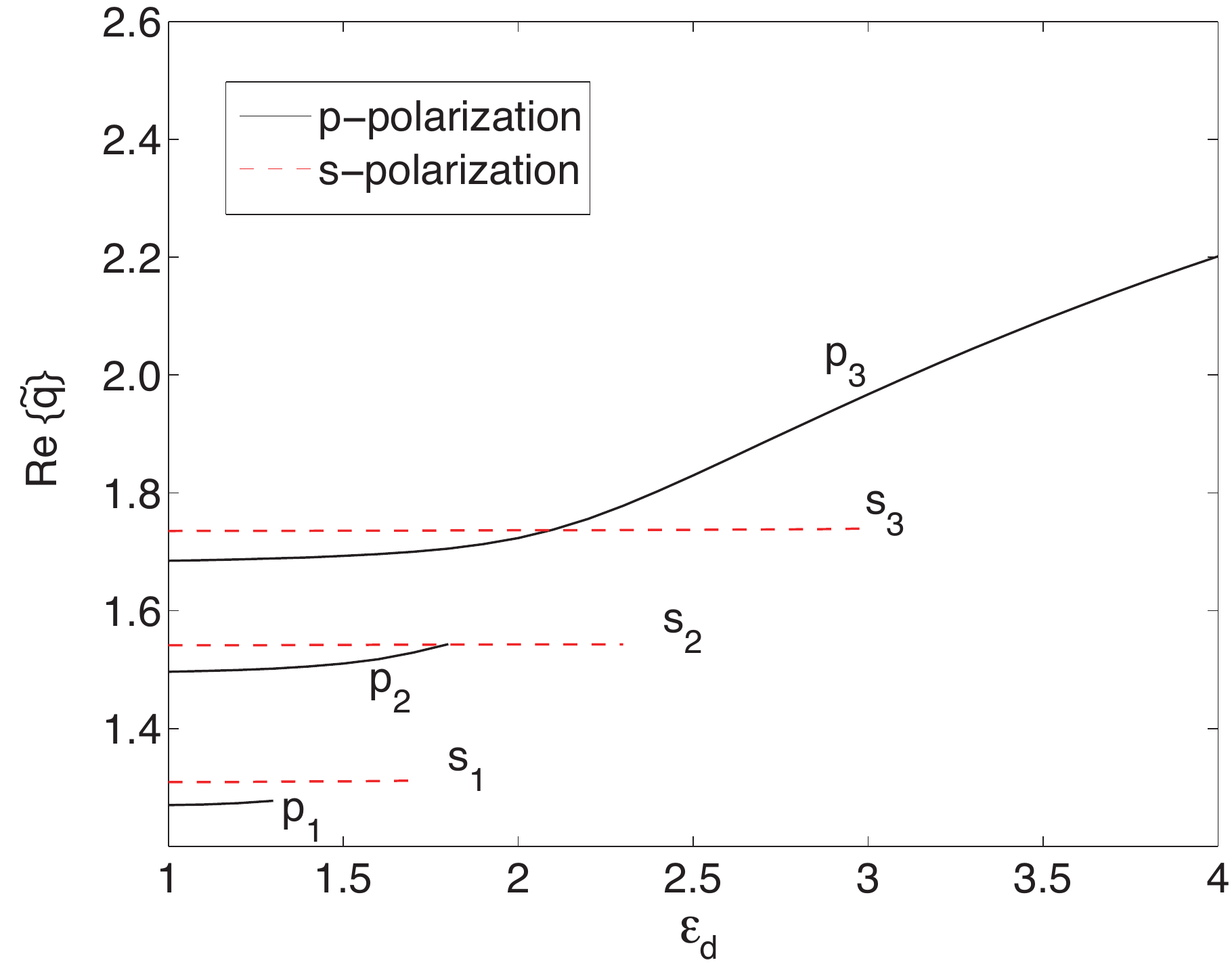}
\end{tabular}
 \end{center}
\caption{Variation of ${\rm Re}(\tq)$ as a function of the relative permittivity $\eps_d$ of the HID material  when $L=25$~nm.  
\label{fig:Req}}
 \end{figure}

\begin{figure}
 \begin{center}
 \begin{tabular}{c}
(a) \includegraphics[width=0.35\linewidth]{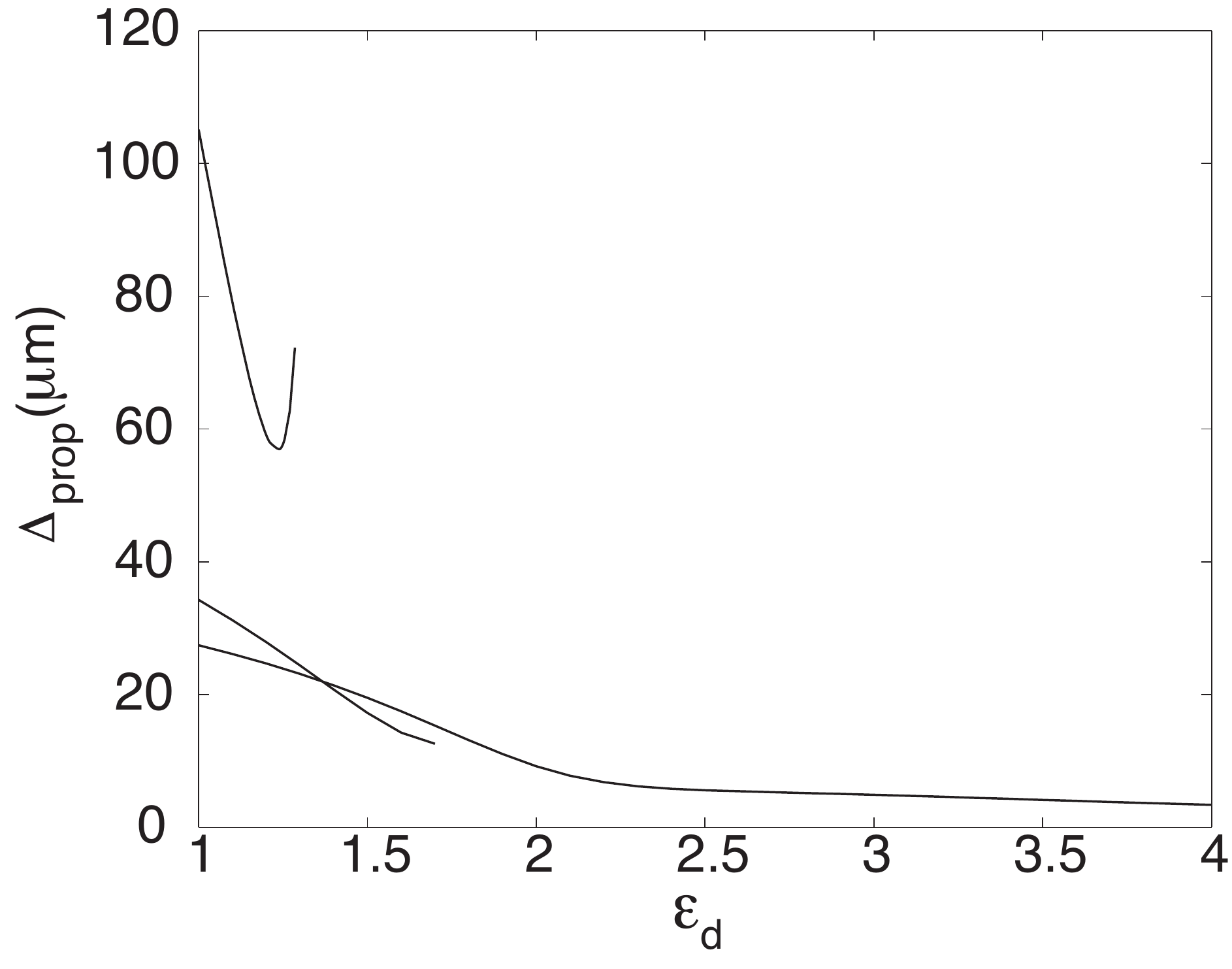}\hspace{10pt}
(b)  \includegraphics[width=0.35\linewidth]{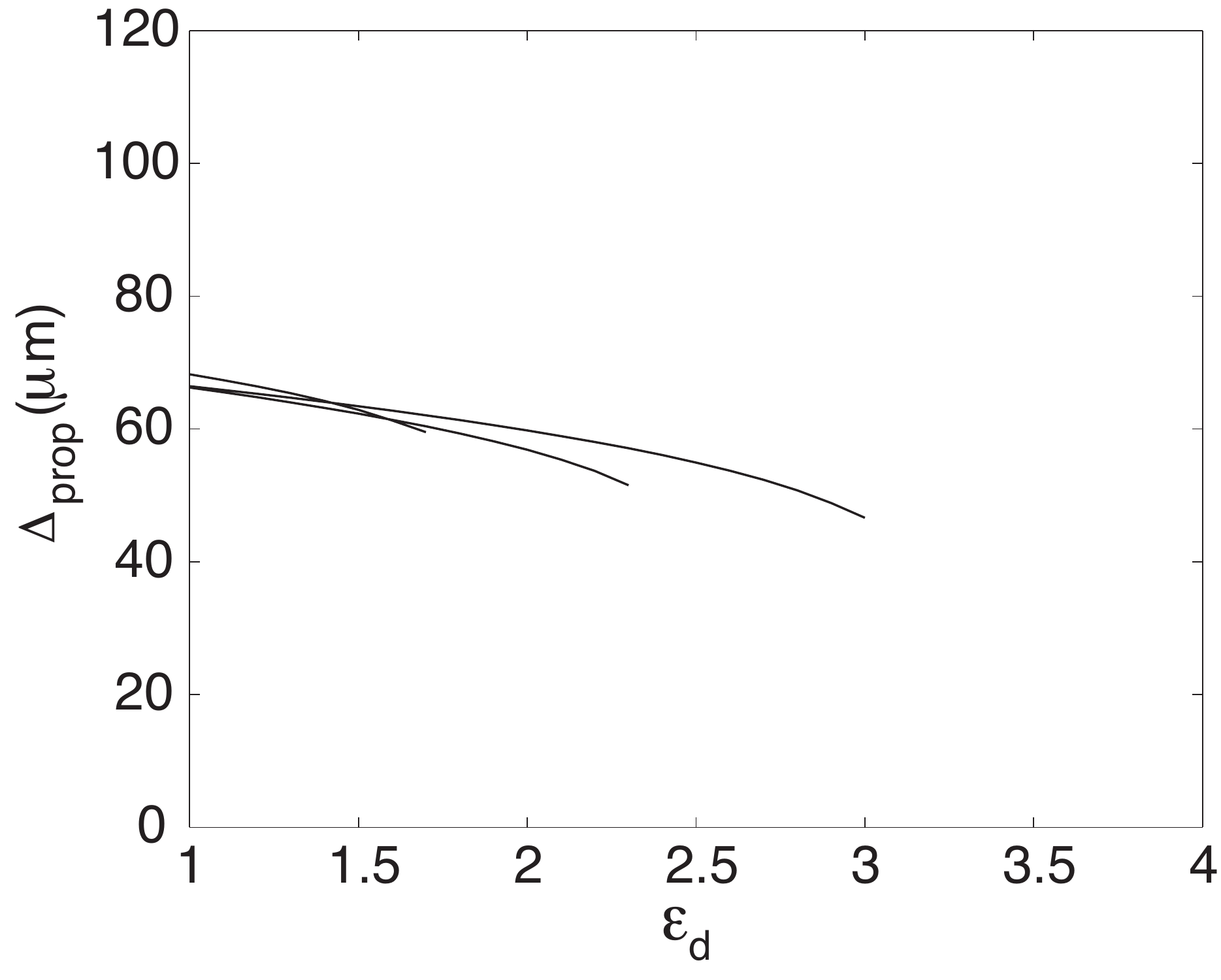}
   \end{tabular}
 \end{center}
\caption{Variation of the propagation distance $\propdist$ as a function of the relative permittivity
$\eps_d$ of the HID material  when $L=25$ nm. (a) $p$-polarized compound SPP waves, (b) $s$-polarized compound SPP waves. 
 \label{fig:Deltaprop}}
 \end{figure}

Figures~\ref{fig:Req}, and \ref{fig:Deltaprop}, respectively, provide plots of ${\rm Re}(\tq)$ and $\propdist$ as functions of
$\eps_d$. A multiplicity of compound SPP waves is evident. We have organized these waves in sets labeled $p_1$ to $p_3$
and $s_1$ to $s_3$, according to the polarization state.
Six compound SPP waves, three $p$ polarized and three $s$ polarized, propagate for $ \eps_d \lesssim 1.3 $. 
These waves begin to disappear one by one ss $ \eps_d $  increases,  reducing to a sole $p$-polarized wave for 
$ \eps_d \gtrsim 3$. As $ \eps_d $ increases from unity, the phase speed $ \vph $ is practically constant  for all $s$-polarized waves as well as in set  $ p_1 $; $\vph$ is almost constant in set $p_2$ for $\eps_d\lesssim 1.5$ and then increases linearly with $\eps_d$; and $\vph$ is almost constant in set $p_3$ for $\eps_d\lesssim 2$ and then increases linearly with $\eps_d$.

The propagation distances $\propdist$ of all three $s$-polarized compound SPP waves are almost equal when $\eps_d=1$,
and all decrease almost linearly with the same rate as $\eps_d$ increases. All three $p$-polarized compound SPP waves are different when $\eps_d=1$; whereas $\propdist$   decreases almost linearly in set $ p_2 $ as $\eps_d$
increases, the  sets $ p_1 $ and $ p_3 $ show more complicated dependencies of $\propdist$ on $\eps_d$.

\begin{figure}
  \begin{center}
  \begin{tabular}{c}
 (a) \includegraphics[width=0.35\linewidth]{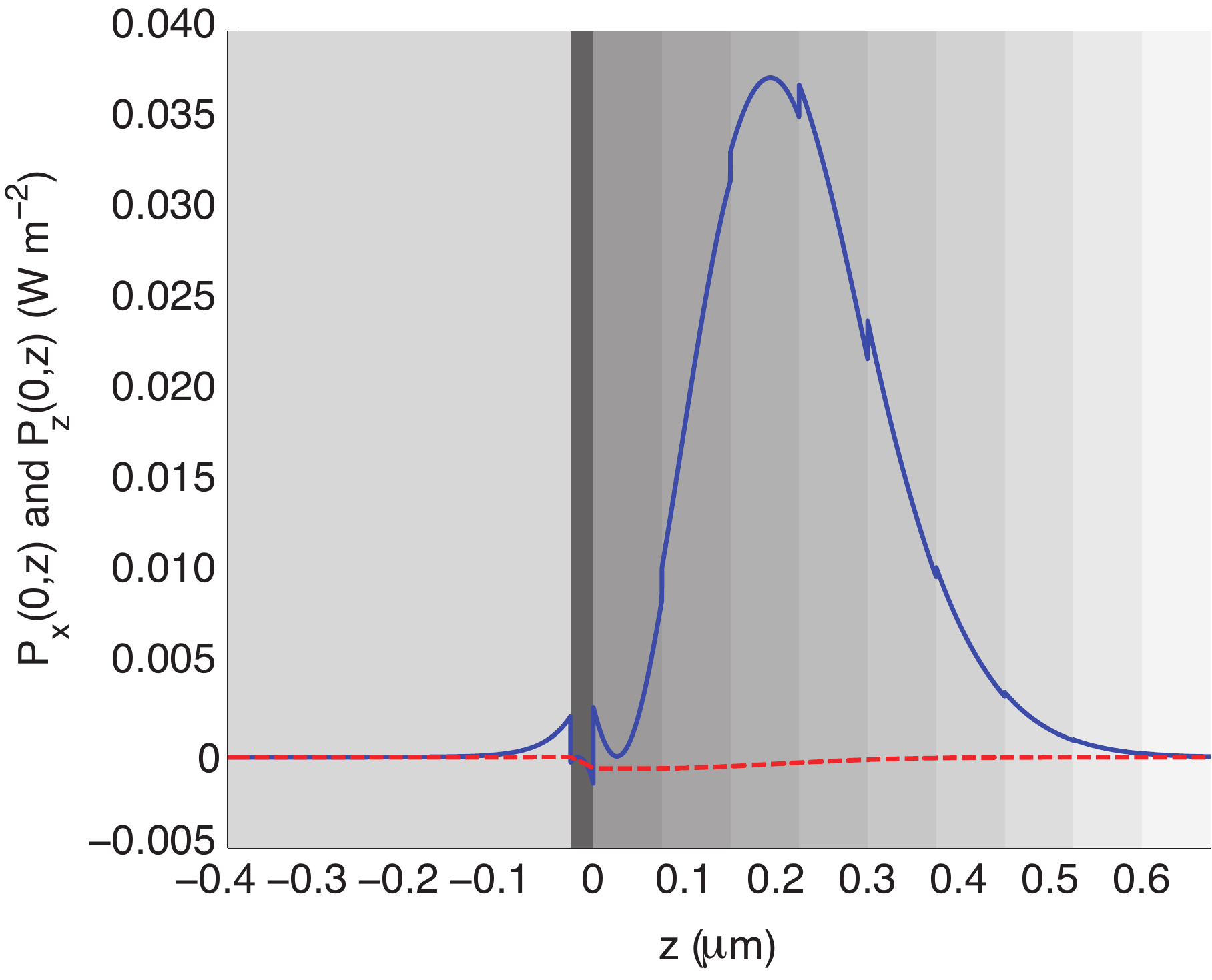}\hspace{10pt}
 (b)  \includegraphics[width=0.35\linewidth]{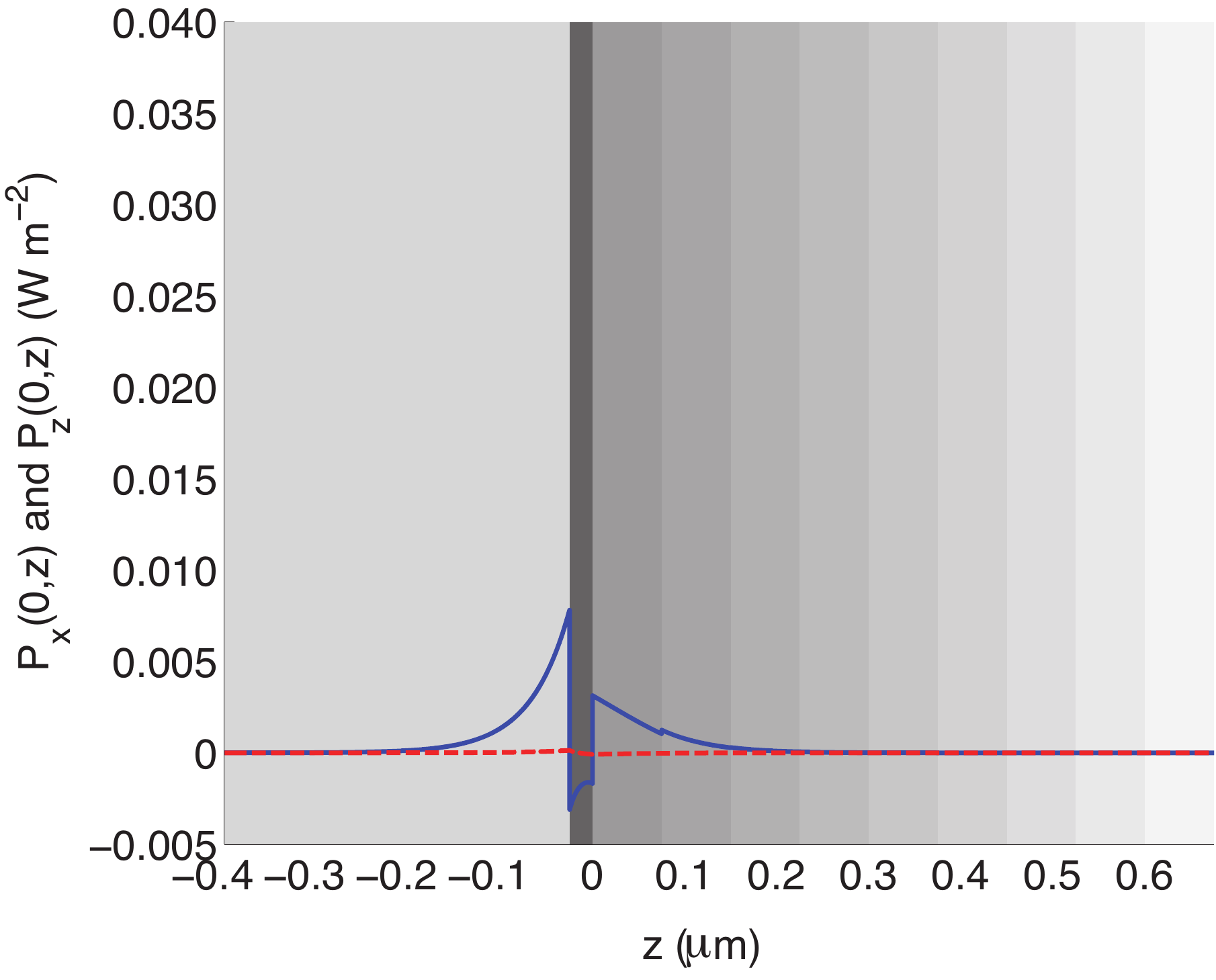}
    \end{tabular}
  \end{center}
 \caption{Spatial variations of $P_x(0,z)$ (blue solid lines) and $P_z(0,z)$ (red dashed lines) with respect to $z$
 for two compound $p$-polarized SPP waves in set $p_3$ when $L=25$~nm and (a) $\eps_d=1$ or (b) $\eps_d=4$.
  \label{fig:CSSPppol}}
  \end{figure}

\begin{figure}
  \begin{center}
  \begin{tabular}{c}
 (a) \includegraphics[width=0.35\linewidth]{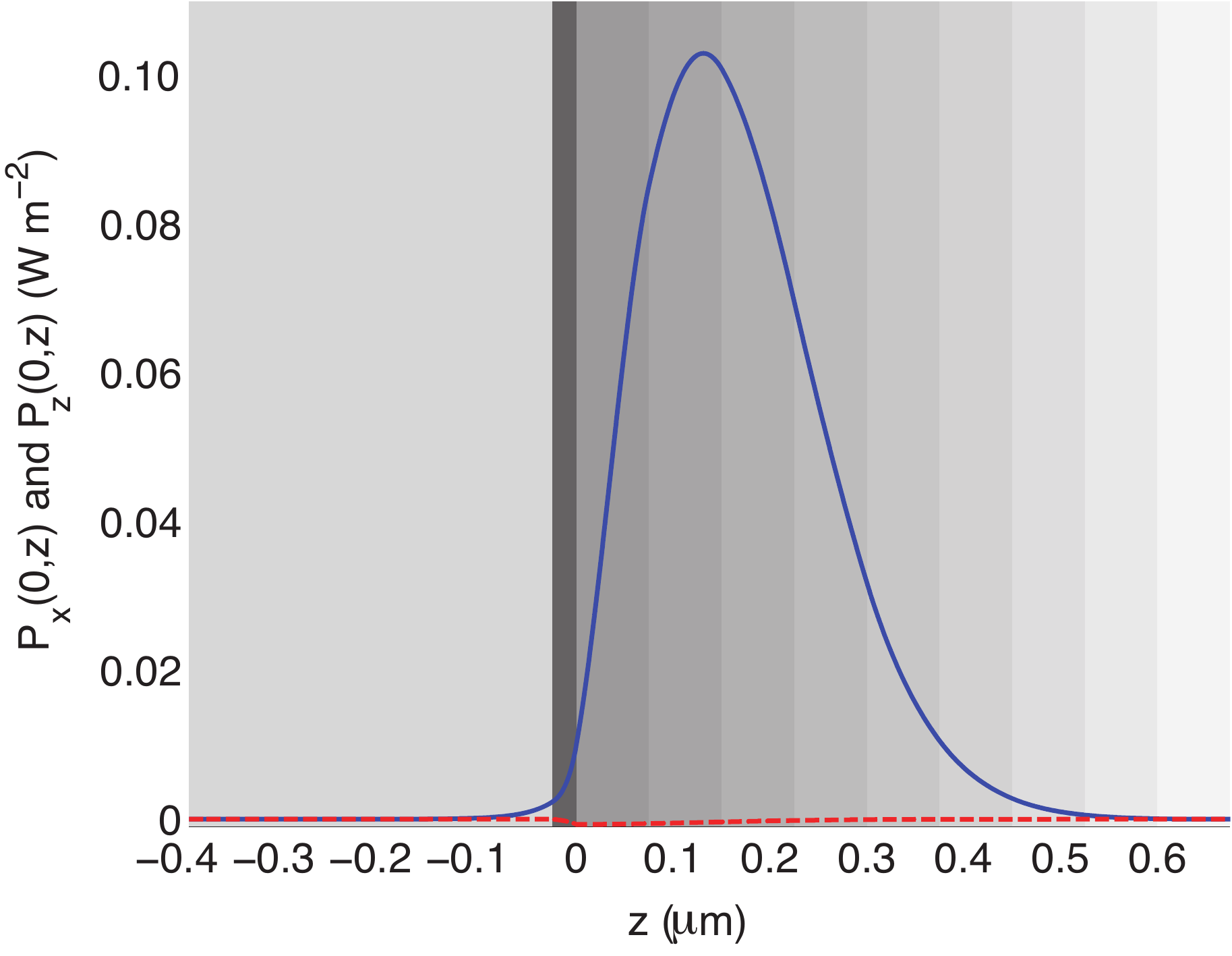}\hspace{10pt}
 (b)  \includegraphics[width=0.35\linewidth]{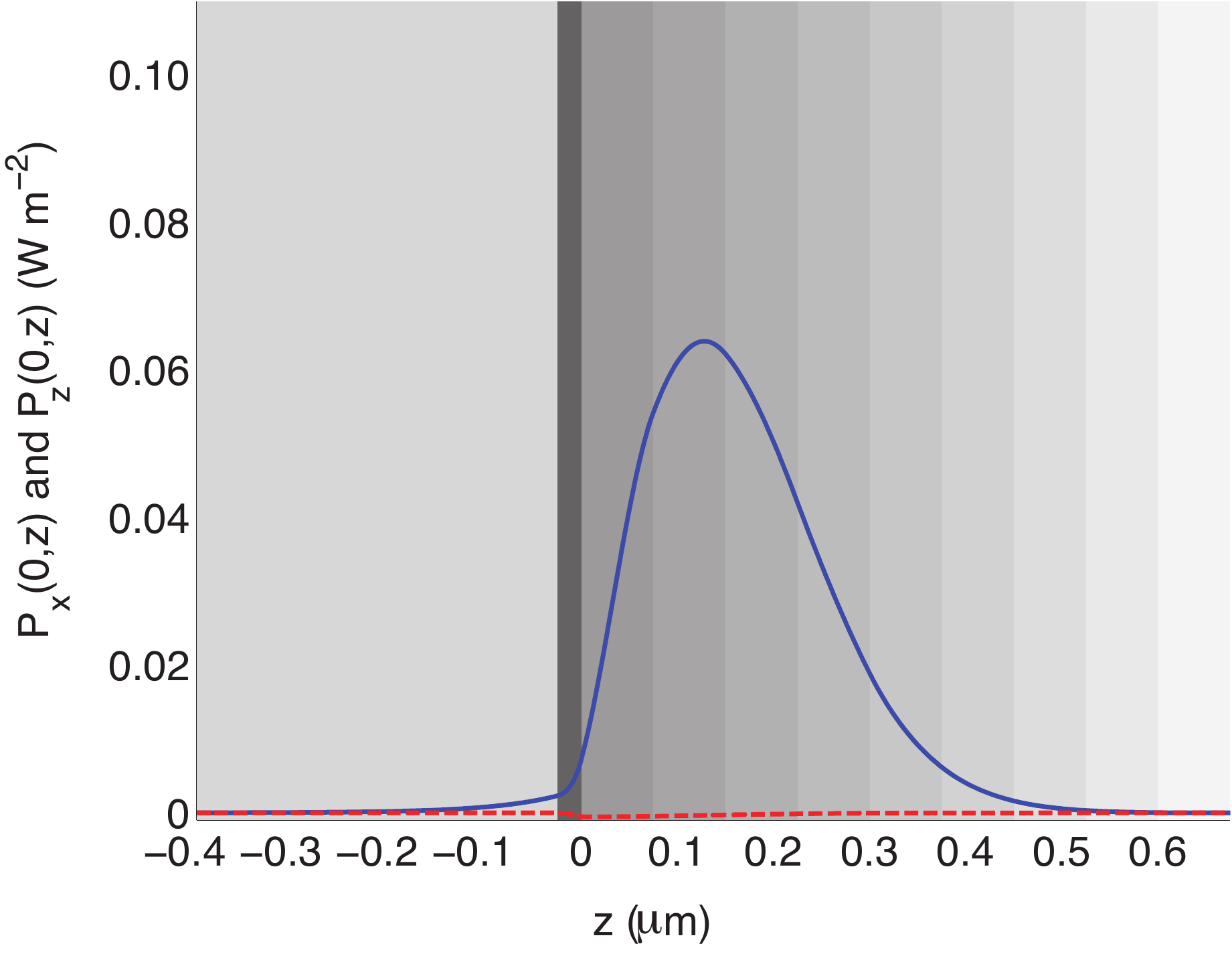}
    \end{tabular}
  \end{center}
 \caption{Spatial variations of $P_x(0,z)$ (blue solid lines) and $P_z(0,z)$ (red dashed lines) with respect to $z$
 for two compound $s$-polarized SPP waves in set $s_3$ when $L=25$~nm and (a) $\eps_d=1$ or (b) $\eps_d=3$.
  \label{fig:CSSPspol}}
  \end{figure}

Finally, Figs.~\ref{fig:CSSPppol} and \ref{fig:CSSPspol} show the spatial profiles of the 
Cartesian components of the time-averaged Poynting vector for two compound SPP waves belonging to the sets $p_3$ and $ s_3 $, respectively, for two different values of $ \eps_d $. Due to the significant coupling between the 
two dielectric/metal interfaces,  a fraction of the energy of all compound SPP waves resides in the HID material and 
that $p$-polarized compound SPP waves are more  sensitive than $s$-polarized compound SPP waves to that material's  relative permittivity.

 \section{Concluding remarks}\label{sec:concl}
 
 We formulated and solved the boundary-value problem for compound surface  waves guided by a thin metal slab
sandwiched between  a homogeneous isotropic dielectric material   and a periodically multilayered isotropic dielectric  material. Solutions were found to exist for compound surface-plasmon-polariton waves of both $p$- and $s$-polarization states at a fixed frequency (or free-space wavelength, equivalently).

For any thickness of the metal layer, numerical investigations indicate that at least one compound surface wave must exist. It possesses the $p$-polarization state, is strongly bound to the metal/HID interface when the metal thickness is large but to both the metal/HID and metal/PMLID interfaces when the metal thickness is small. When the metal layer vanishes, this compound SPP wave transmutes into a Tamm wave. The behavior of this compound SPP wave clearly shows the coalescence of two surface-wave phenomenons that otherwise appear completely different.

Additional compound SPP waves exist, depending on the thickness of the metal layer and the relative permittivity of the
HID material. Some  of these are $p$ polarized, the others being $s$ polarized. All of them differ in phase speed, attenuation rate, and field profile, even though all are excitable at the same frequency. We conjecture that the multiplicity is enhanced when the metal layer is thin if the relative permittivity of the HID material is  smaller than the spatially averaged relative permittivity of the PMLID material.
Although not explored here but extrapolation from previous work on SPP waves guided by a metal/PMLID interface
\cite{FHBML,Liujnp,Hallnano,AkhBook}, we can state that the period and the composition of the PMLID material also play a significant role in the multiplicity of compound SPP waves.
The multiplicity and the dependence of the number of compound SPP waves on the relative permittivity of the HID material when the metal layer is thin could be useful for optical sensing applications.

\end{document}